\documentclass[reprint,amssymb,amsmath,superscriptaddress,aps,showpacs,10pt,floatfix,longbibliography,prl]{revtex4-1}

\usepackage[pdftex]{graphicx} 
\usepackage{epstopdf}
\usepackage{verbatim}
\usepackage{amsmath}
\usepackage{color}
\usepackage{subfigure}
\usepackage{amsbsy}
\usepackage{wasysym}

\begin{document}

\title{$\mu$SR evidence for the U(1) quantum spin liquid ground state\\ in the triangular antiferromagnet YbMgGaO$_4$}

\author{Yuesheng Li}
\email{yuesheng.man.li@gmail.com}
\affiliation{Department of Physics,
Renmin University of China,
Beijing 100872, P. R. China}
\affiliation{Experimental Physics VI, Center for Electronic Correlations and Magnetism, University of Augsburg, 86159 Augsburg, Germany}

\author{Devashibhai Adroja}
\affiliation{ISIS Pulsed Neutron and Muon Source, STFC Rutherford Appleton Laboratory, Harwell Campus, Didcot, Oxfordshire, OX11 0QX, United Kingdom}
\affiliation{Highly Correlated Matter Research Group, Physics Department,University of Johannesburg, PO Box 524, Auckland Park 2006, South Africa}

\author{Pabitra K. Biswas}
\affiliation{ISIS Pulsed Neutron and Muon Source, STFC Rutherford Appleton Laboratory, Harwell Campus, Didcot, Oxfordshire, OX11 0QX, United Kingdom}

\author{Peter J. Baker}
\affiliation{ISIS Pulsed Neutron and Muon Source, STFC Rutherford Appleton Laboratory, Harwell Campus, Didcot, Oxfordshire, OX11 0QX, United Kingdom}

\author{Qian Zhang}
\affiliation{Department of Physics,
Renmin University of China,
Beijing 100872, P. R. China}

\author{Juanjuan Liu}
\affiliation{Department of Physics,
Renmin University of China,
Beijing 100872, P. R. China}

\author{Alexander A. Tsirlin}
\affiliation{Experimental Physics VI, Center for Electronic Correlations and Magnetism, University of Augsburg, 86159 Augsburg, Germany}

\author{Philipp Gegenwart}
\affiliation{Experimental Physics VI, Center for Electronic Correlations and Magnetism, University of Augsburg, 86159 Augsburg, Germany}

\author{Qingming Zhang}
\email{qmzhang@ruc.edu.cn}
\affiliation{Department of Physics,
Renmin University of China, Beijing 100872, P. R. China}
\affiliation{Collaborative Innovation Center of Advanced Microstructures, Nanjing 210093, P. R. China}

\date{\today}

\begin{abstract}
Muon spin relaxation and rotation ($\mu$SR) experiments on single crystals of the structurally perfect triangular antiferromagnet YbMgGaO$_4$ indicate the absence of both static long-range magnetic order and spin freezing down to 0.048\,K in zero field. Below 0.4\,K, the $\mu^{+}$ spin relaxation rates, which are proportional to the dynamic correlation function of the Yb$^{3+}$ spins, 
exhibit temperature-independent plateaus. Same behavior is revealed by temperature-independent local susceptibilities extracted from the Knight shifts of the $\mu^{+}$ spin rotation frequencies under a transverse field of 20\,Oe. All these $\mu$SR results unequivocally support the formation of a gapless U(1) quantum spin liquid ground state in the triangular antiferromagnet YbMgGaO$_4$.
\end{abstract}


\maketitle

\emph{Introduction.}---Antiferromagnetically coupled ($J>0$) spins on a perfect geometrically frustrated lattice, such as the triangular or kagome lattices, can preserve strong fluctuations and evade long-range order or spin freezing even at $T\ll J$. They reveal exotic phases characterized by interesting properties, such as fractionalized spin excitations, intrinsic topological order and gapless excitations without symmetry breaking. These new phases have been proposed as quantum spin liquids (QSL)~\mbox{\cite{balents2010spin,lee2008end,wen2004quantum}}. The spin-$\frac12$ triangular Heisenberg antiferromagnet, initially believed to host a resonating-valence-bond QSL ground state (GS)~\cite{anderson1973resonating,anderson1987resonating,baskaran1987resonating}, develops in fact the $120^{\circ}$-type magnetic order~\cite{singh1992three,sindzingre1994investigation,bernu1994exact,capriotti1999long,weihong1999phase}. However, this order is very fragile and can be melted by perturbations, such as next-nearest-neighbor couplings~\cite{kaneko2014gapless,li2015quasiclassical}, spatially anisotropic interactions~\cite{trumper1999spin,yunoki2006two,ohashi2008finite}, and bond randomness~\cite{watanabe2014quantum}. Theoretical studies found that ring couplings destroy long-range order as well, and trigger the formation of a U(1) QSL GS with a spinon Fermi surface~\cite{motrunich2005variational,lee2005u,motrunich2006orbital}. On the other hand, relevant experimental systems, $\kappa$-(BEDT-TTF)$_2$Cu$_2$(CN)$_3$~\cite{yamashita2008thermodynamic} and EtMe$_3$Sb[Pd(dmit)$_2$]$_2$~\cite{yamashita2011gapless}, revealed linear temperature dependence of the heat capacity, in contrast to $C_v\sim T^{2/3}$ predicted for the U(1) QSL~\cite{motrunich2005variational}.

\begin{figure}[t]
\begin{center}
\includegraphics[width=8.5cm,angle=0]{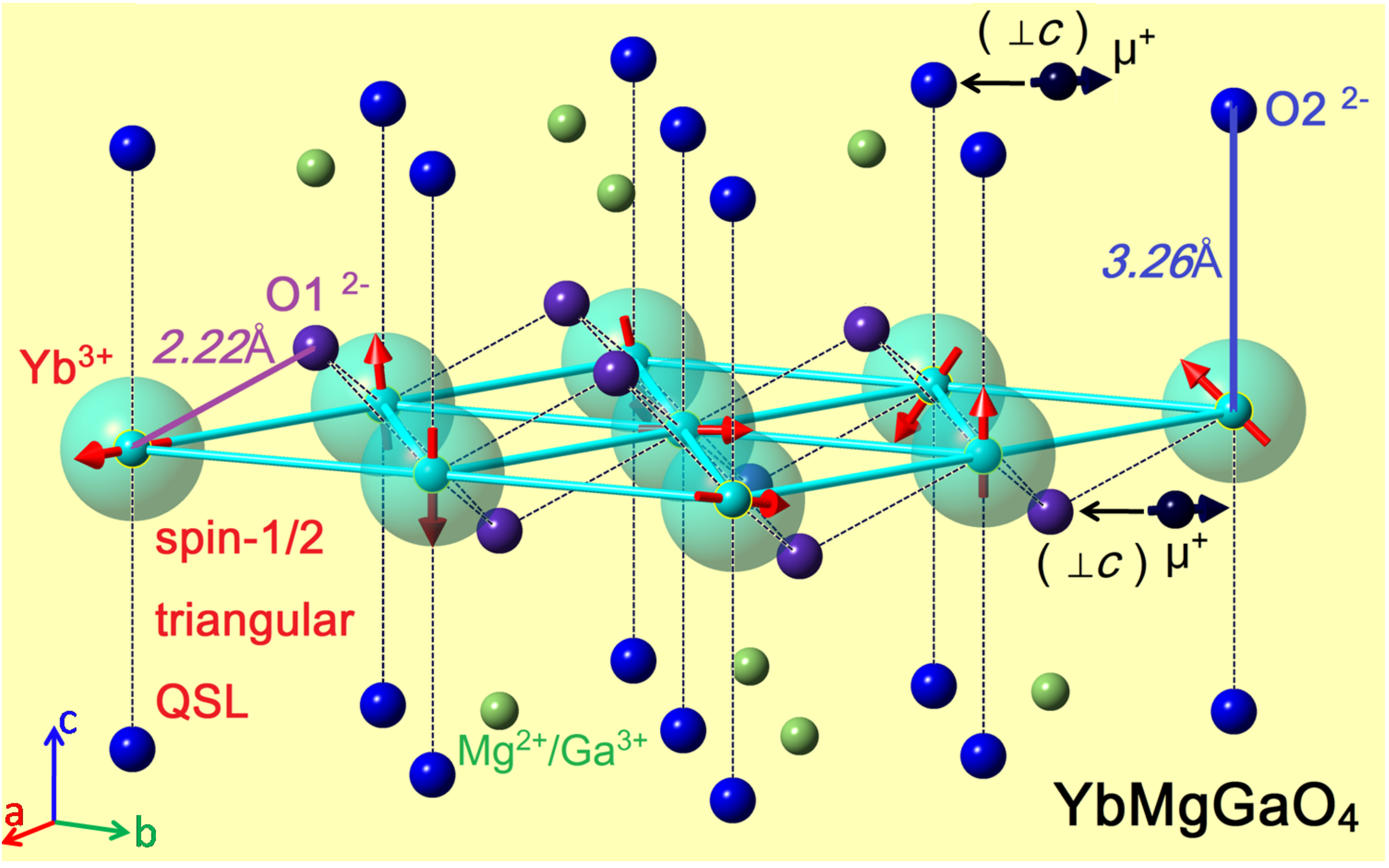}
\caption{(Color online)
Crystal structure of YbMgGaO$_4$ around the effective spin-$\frac12$ triangular layer of Yb$^{3+}$. The implanted $\mu^+$ is likely to stop near O1$^{2-}$ and O2$^{2-}$, and probes the local spin dynamics of the triangular QSL.}
\label{fig1}
\end{center}
\end{figure}

Recently, our group reported a new structurally perfect rare-earth triangular antiferromagnet YbMgGaO$_4$~\mbox{\cite{li2015gapless,li2015rare}}. Unlike the majority of QSL candidates, YbMgGaO$_4$ is free from magnetic defects~\cite{lee2007quantum,freedman2010site,li2012structure,li2013transition,li2014gapless}, spatial anisotropy~\cite{li2014gapless,shimizu2003spin,itou2008quantum}, and antisymmetric Dzyaloshinsky-Moriya anisotropy~\cite{moriya1960new,zorko2008dzyaloshinsky}. Its magnetic heat capacity reveals the $C_v\sim T^{2/3}$ behavior with almost zero residual entropy down to $\sim$ 0.06 K~\cite{li2015gapless} compatible with the triangular U(1) QSL GS~\cite{motrunich2005variational}. The spin susceptibility of a U(1) QSL is expected to approach a constant value as the temperature goes down to zero~\cite{motrunich2005variational,lee2005u,motrunich2006orbital}. While the divergent nature of the bulk static susceptibility of YbMgGaO$_4$ was measured down to 0.48 K~\cite{li2015gapless}, this experimental temperature range was certainly not low enough to probe GS properties in a system, where $J_0$ is as low as 1.5\,K~\cite{li2015rare}. In the following, we fill this gap by probing YbMgGaO$_4$ down to 0.048\,K, and provide evidence for the U(1) QSL GS.

Weak magnetic couplings between the rare-earth Yb$^{3+}$ spins render experimental probe of the GS extremely challenging. While nuclear magnetic resonance requires an external field on the order of 1\,T that inevitably perturbs such a system, inelastic neutron scattering can be performed in zero field, but fails to detect inelastic signal at transfer energies below 0.1\,meV, owing to the contamination by the elastic signal. In this respect, muon spin relaxation and rotation ($\mu$SR) is an ideal technique that can be performed in true zero field (ZF). ZF-$\mu$SR is an extremely sensitive probe detecting tiny internal fields on the order of 0.1\,G. The $\mu$SR time window can measure magnetic fluctuation rates in the range from $10^4$ to $10^{12}$\,Hz~\cite{bramwell2009measurement,pratt2011magnetic}.

In this Letter, we report a comprehensive $\mu$SR investigation of the GS spin dynamics of YbMgGaO$_4$ using single crystals. Neither the oscillation signal nor characteristic recovery of the polarization to 1/3 is observed in the ZF measurements with the incident $\mu^{+}$ polarization perpendicular and parallel to the \emph{c}-axis down to 0.048\,K and 0.066\,K, respectively, thus ruling out any static uniform or random field exceeding 0.9 Oe. Both the Knight shift of the $\mu^{+}$ spin rotation frequency ($\propto$ local spin susceptibility) measured under a transverse field (TF) of 20\,Oe and the $\mu^{+}$ spin relaxation rate ($\propto$ spin dynamic correlation function) measured in ZF/TF exhibit a plateau below 0.4\,K. These observations strongly suggest that a gapless U(1) QSL GS is formed in YbMgGaO$_4$.

\emph{Experimental technique.}---Large single crystals ($\sim$ 1 cm)
of YbMgGaO$_4$ were grown by the floating zone technique~\cite{li2015rare}. The high quality of the single crystals was confirmed by X-ray diffraction showing narrow reflections with $\Delta$(2$\theta$) $\sim 0.04^{\circ}$ ~\cite{supple}. The crystal orientations were determined by Laue x-ray diffraction. The crystals were cut into slices along both \emph{c}-axis and \emph{ab}-plane with a homogeneous thickness of $\sim$ 1 mm. Mosaics of slices along the \emph{c}-axis (S1) and the \emph{ab}-plane (S2) were mounted on two silver sample holders~\cite{supple}. The $\mu$SR data were collected at the ISIS pulsed muon facility, Rutherford Appleton Laboratory, U.K on both samples (S1 and S2) between 0.05\,K and 4\,K using dilution refrigerators. Additional data between 2\,K and 50\,K were collected by transferring the sample to a $^4$He cryostat~\cite{hillier2005musr}. The SI units are used, and $\langle$$\rangle$ represents a thermal and sample average.

\begin{figure}[t]
\begin{center}
\includegraphics[width=8.7cm,angle=0]{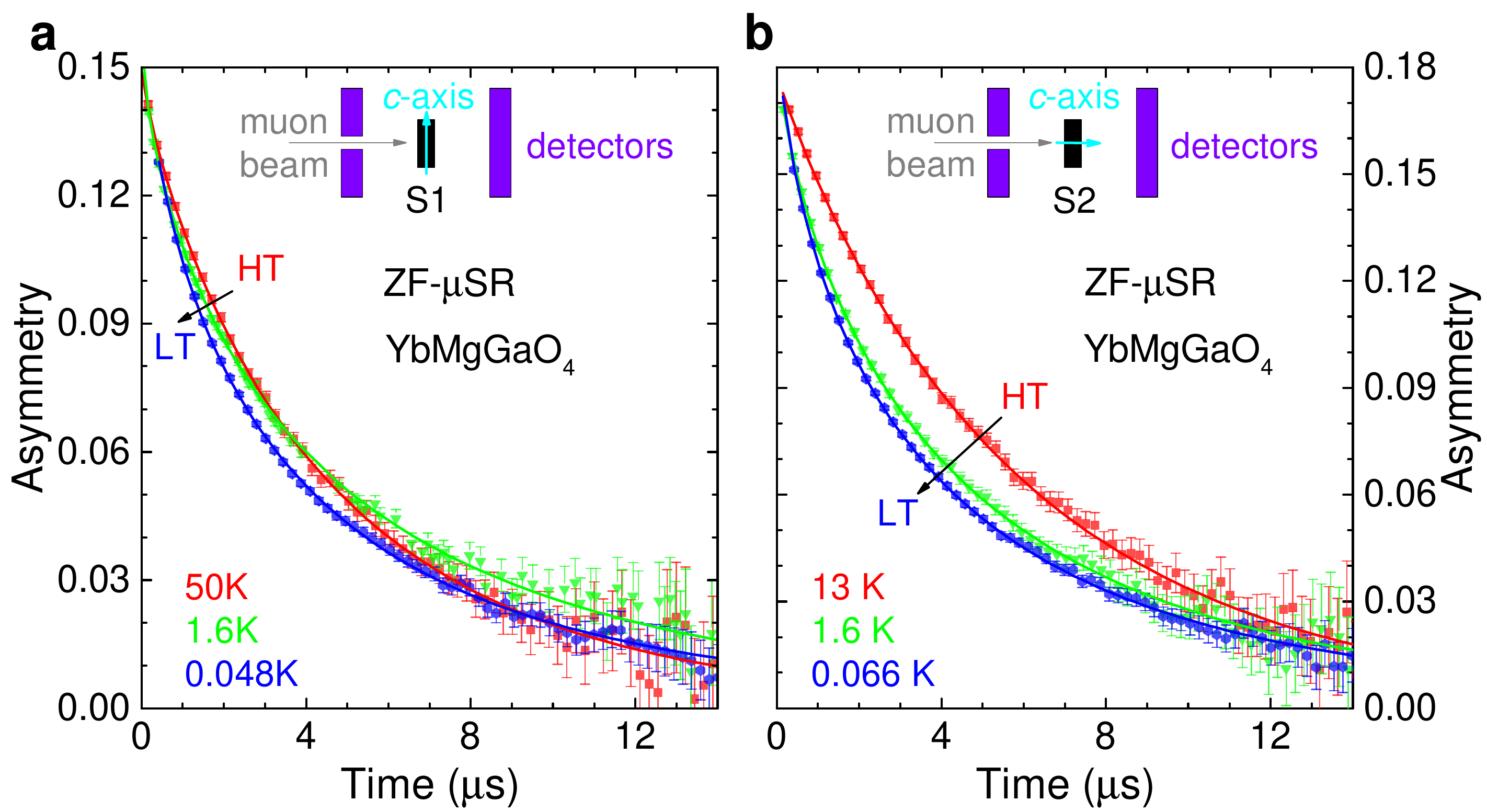}
\caption{(Color online)
Selected ZF-$\mu$SR signals (asymmetry vs time, background-subtracted) with the incident beam a. perpendicular ($\perp$) and b. parallel ($\parallel$) to the \emph{c}-axis. The colored lines are the corresponding fits to the data using Eq.~\eqref{function1}. The insets show relevant experimental geometries.}
\label{fig2}
\end{center}
\end{figure}

\emph{Absence of spin freezing.}---Implanted muons are very sensitive to local magnetic fields induced by the neighboring Yb$^{3+}$ spins (Fig.~\ref{fig1})~\cite{supple}. Therefore, ZF-$\mu$SR is the best tool to detect long-range magnetic order or spin freezing. 

\begin{figure}[t]
\begin{center}
\includegraphics[width=8.5cm,angle=0]{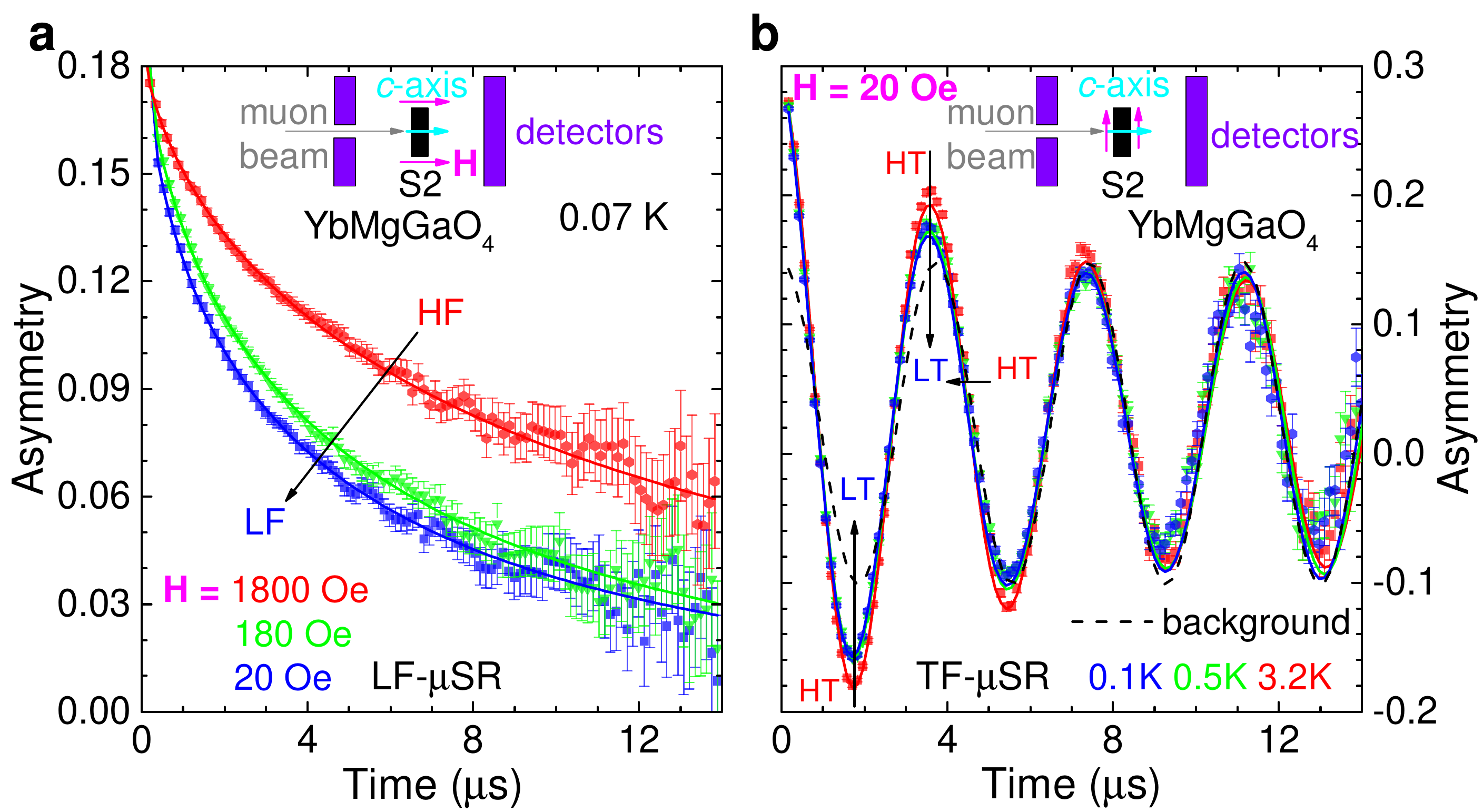}
\caption{(Color online)
a. Selected LF-$\mu$SR signals (background-subtracted) at 0.07 K. The colored lines are fits to the data using Eq.~\eqref{function1}. b. Selected raw TF-$\mu$SR signals measured in the TF field of 20\,Oe. The colored lines are fits to the data using Eq.~\eqref{function2}. The insets show the experimental geometries.}
\label{fig3}
\end{center}
\end{figure}

Our ZF and longitudinal-field (LF) data (Fig.~\ref{fig2} and ~\ref{fig3}a) are well fitted to a stretched exponential relaxation function:
\begin{equation}
A^{\rm ZF/LF} = A_0\exp[-(\lambda t)^{\beta}]+B_{\rm ZF}.
\label{function1}
\end{equation}
The raw TF data (Fig.~\ref{fig3}b) are well fitted to a stretched exponential relaxation function with damped oscillations~\cite{hillier2005musr,keren2004dynamic}:
\begin{equation}
A^{\rm TF} = A_0\exp[-(\lambda t)^{\beta}]\cos(\omega t)+B_{g}\cos(\omega_0 t)+B_{\rm TF}.
\label{function2}
\end{equation}
Here, $A_0\sim 0.2$ is the initial asymmetry (weakly temperature- and field-dependent), $\lambda$ is the $\mu^+$ spin relaxation rate, and $\beta$ is the stretching exponent. The coefficients $B_{\rm ZF}\sim 0.13$, $B_{g}\sim 0.12$, and $B_{\rm TF}\sim 0.02$ are background constants representing muons that missed the sample ($B_{\rm ZF}$), missed the sample but stayed in the TF or stopped in the Ag-holder ($B_{g}$), and missed both the sample and the TF ($B_{\rm TF}$). The frequencies $\omega=2\pi\gamma_{\mu}B_{\rm loc}$ and $\omega_0=2\pi\gamma_{\mu}B_{0}$ are caused by the local field $B_{\rm loc}$ and by the external TF $B_{0}$, respectively, and $\gamma_{\mu}=135.5$\,MHz/T is the $\mu^+$ gyromagnetic ratio.

The absence of spin freezing in YbMgGaO$_4$ is supported by the following observations:

\romannumeral1) The ZF asymmetries decrease continuously (see Fig.~\ref{fig2}) without showing oscillations in the full time window, $0-20$\,$\mu$s. This continuous decrease is observed down to 0.048\,K and 0.066\,K with the incident $\mu^+$ polarization perpendicular and parallel to the \emph{c}-axis, respectively~\cite{supple}, suggesting that no static uniform local field is formed. 

\romannumeral2) The ZF relaxations also lack a recovery of the polarization to 1/3, suggesting the absence of static random fields (no spin-glass-like freezing)~\cite{uemura1985muon,uemura1994spin}. 

\begin{figure}[t]
\begin{center}
\includegraphics[width=8.5cm,angle=0]{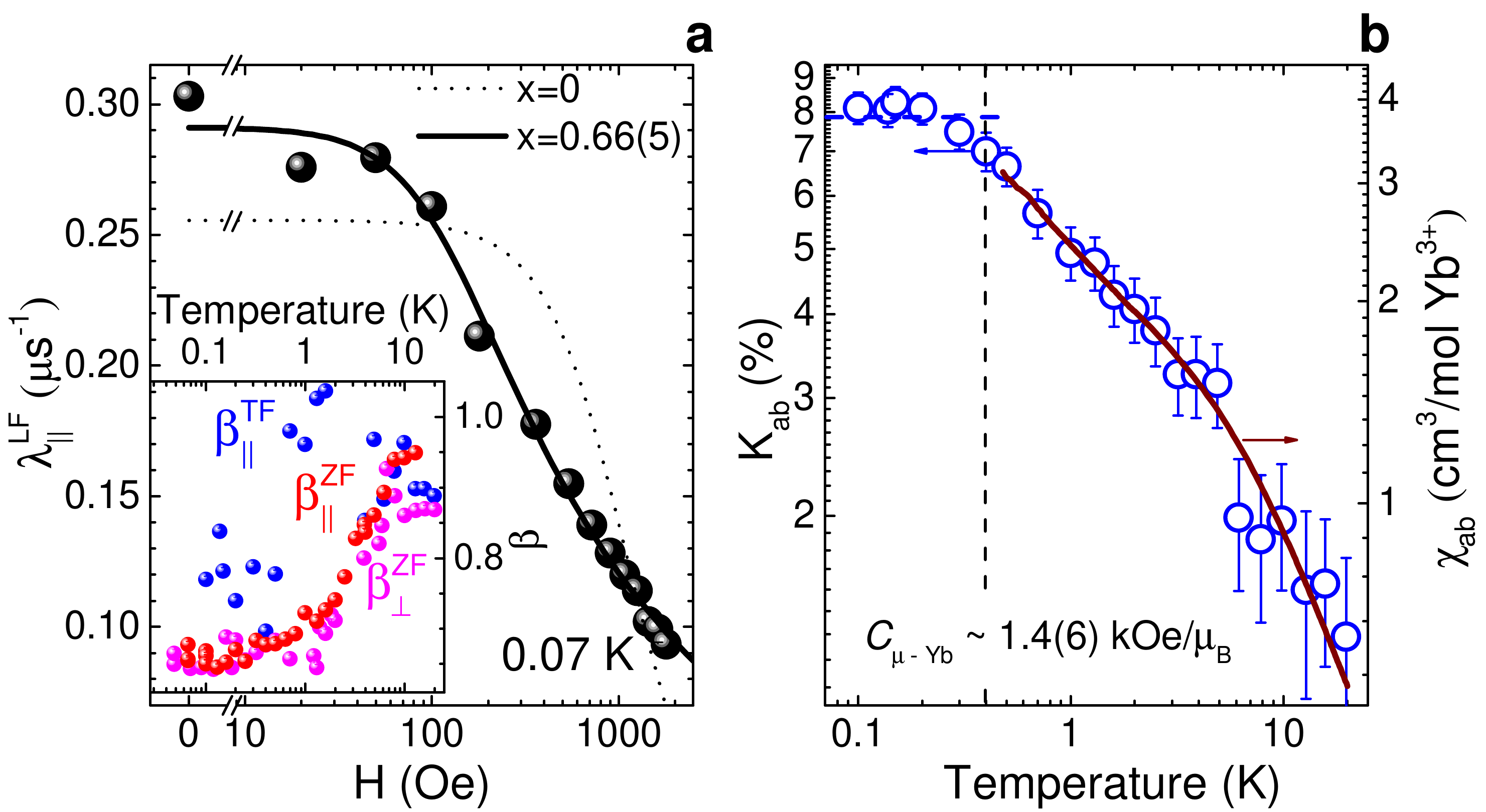}
\caption{(Color online)
a. LF dependence of the $\mu^+$ spin relaxation rate, $\lambda_{\parallel}^{\rm LF}$(H), at 0.07\,K. The dotted line represents the fit to the $\lambda_{\parallel}^{\rm LF}$ using Eq.~\eqref{relaxation rate1} ($x=0$), and the solid line is the fit using Eq.~\eqref{relaxation rate2} with $x=0.66(5)$. Inset: temperature dependence of the stretching exponents, $\beta$. b. Temperature dependence of the Knight shift obtained from the TF-$\mu$SR measurements and from the bulk static susceptibility measured under 100\,Oe down to 0.48\,K in the field applied along the $ab$-plane.}
\label{fig4}
\end{center}
\end{figure}

\romannumeral3) The stretching exponent $\beta$ gradually decreases from $\sim$ 1 at high temperatures ($T\geq 4$\,K $\gg J_{0}$) down to a constant value of $\sim$ 0.6 at the lowest temperatures ($T\sim 0.1$\,K $\ll J_{0}$), see the inset of Fig.~\ref{fig4}a. In contrast, for a spin glass one expects that $\beta$ decreases from 1 at high temperatures to 1/3 at the spin freezing temperature~\mbox{\cite{ogielski1985dynamics,campbell1994dynamics}}. 

\begin{figure}[t]
\begin{center}
\includegraphics[width=8.5cm,angle=0]{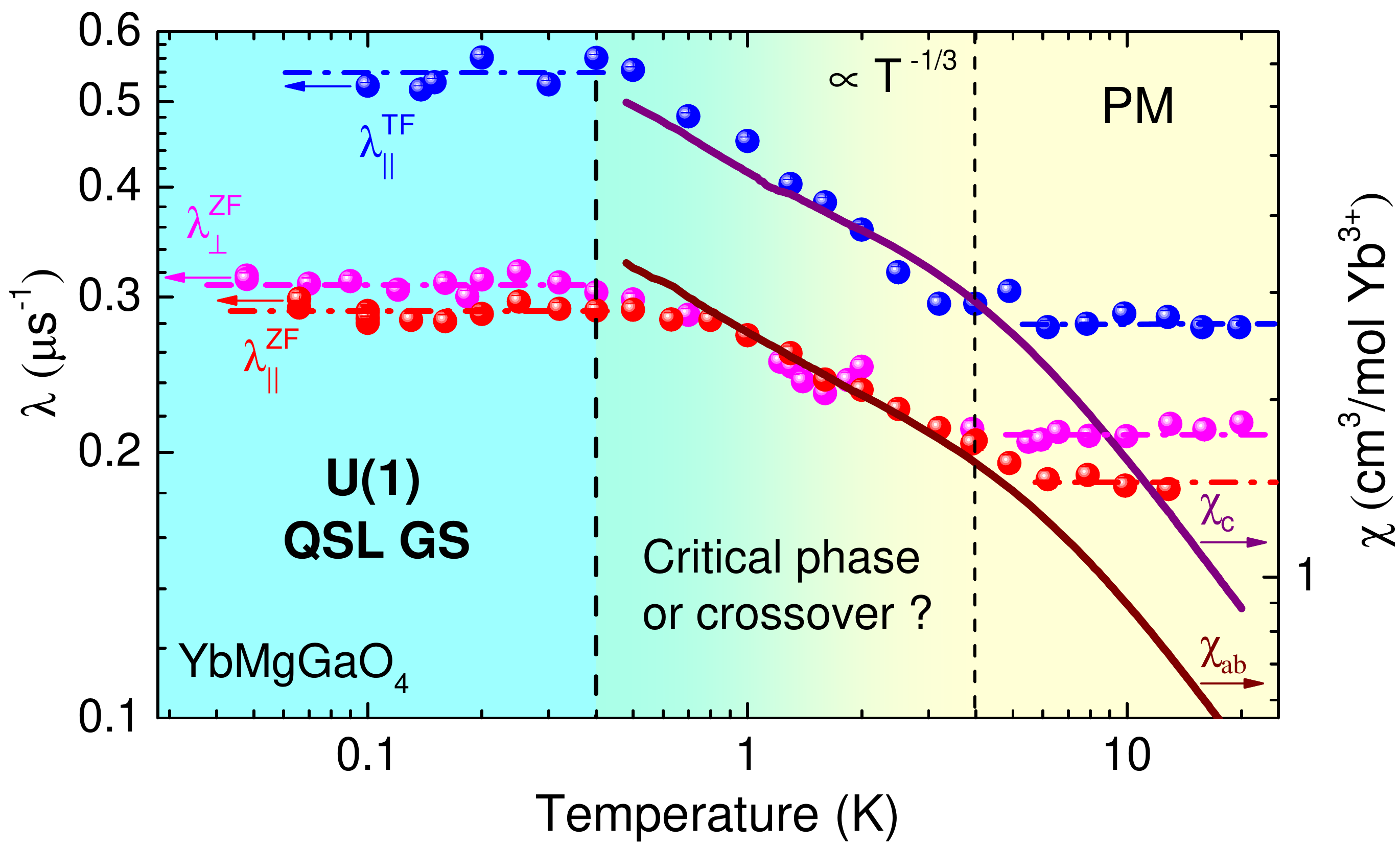}
\caption{(Color online)
Temperature dependence of the $\mu^{+}$ spin relaxation rates from the ZF and TF measurements and from the bulk static susceptibilities measured in the applied field of 100\,Oe.}
\label{fig5}
\end{center}
\end{figure}

\romannumeral4) The ZF $\mu^+$ spin relaxation rate $\lambda^{\rm ZF}$ increases by 50\,\% upon cooling from $T\gg J_{0}$ ($\lambda^{\rm ZF}\sim 0.2$\,$\mu$s$^{-1}$) down to $T\ll J_{0}$ ($\lambda^{\rm ZF}\sim 0.3$\,$\mu$s$^{-1}$), see Fig.~\ref{fig2} and Fig.~\ref{fig5}. This indicates only a weakly slowing down of the Yb$^{3+}$ spin fluctuations~\cite{keren2004dynamic,keren2001probing,bono2004mu}, whereas in a spin glass the relaxation rate will typically increase by several orders of magnitude below the freezing point ($\lambda\sim 1-20$\,$\mu$s$^{-1}$)~\cite{uemura1985muon,uemura1994spin,bono2004mu}. 

Our observations indicate that the relatively strong phase-coherent quantum fluctuations of Yb$^{3+}$ spins survive at low temperatures. Upon changing the temperature from above 4\,K to below 0.4\,K and suppressing the energy of thermal fluctuations by at least one order of magnitude, about 2/3 of the total fluctuations at $T>4$\,K are retained due to the strong geometrical frustration~\cite{balents2010spin}. The spin relaxation rate $\lambda^{\rm ZF}$ saturates below $T_s\simeq 0.4$\,K (Fig.~\ref{fig5}). This characteristic temperature of 0.4\,K is nearly the same as in other QSL candidates~\cite{kermarrec2011spin,faak2012kapellasite,clark2013gapless,gomilsek2016mu,gomilvsek2016instabilities} and, remarkably, quite high on the relative energy scale of YbMgGaO$_4$ ($T_s/J_0\sim 0.27$). We suggest that quantum fluctuations have paramount effect below this temperature, regardless of the absolute scale of the exchange couplings $J_0$.

The upper limit of local static uniform or random fields can be estimated as ($\langle B_{\rm loc}^{ab}\rangle+\langle B_{\rm loc}^{c}\rangle)/2<0.9$\,Oe down to 0.048\,K and $\langle B_{\rm loc}^{ab}\rangle<0.9$\,Oe down to 0.066\,K~\cite{supple}. 

\emph{Knight shift.}---The TF-$\mu$SR time spectra are shown in Fig.~\ref{fig3}b for the TF of 20 Oe with the incident $\mu^+$ polarization parallel to the \emph{c}-axis. The frequency of the muon spin rotation is proportional to the local field in the \emph{ab}-plane, $\omega=\gamma_{\mu}B_{\rm loc}^{ab}$, and the Knight shift should be proportional to the static $\mathbf q=0$ susceptibility $\chi$,
\begin{equation}
K_{ab} = \frac{\omega-\omega_0}{\omega_0} = \frac{C_{\mu-{\rm Yb}}\times\chi_{ab}}{4\pi N_{A}\mu_{B}},
\label{knight shift}
\end{equation}
where $C_{\mu-{\rm Yb}}\sim 1.4(6)$\,kOe/$\mu_B$ represents the effective dipole magnetic coupling constant between the implanted $\mu^+$ spins and the Yb$^{3+}$ spins~\cite{gomilsek2016mu}. $N_{A}$ and $\mu_{B}$ are Avogadro's constant and Bohr magneton, respectively. As the temperature goes down, the Knight shift follows the bulk static susceptibility down to 0.48\,K (see Fig.~\ref{fig4}b), which is the lowest temperature of the susceptibility measurement. Below 0.4\,K, the Knight shift saturates at a constant value suggesting the zero-temperature spin susceptibility of $\sim$ 3.5 cm$^3$/mol Yb$^{3+}$, which is consistent with the U(1) QSL GS with a spinon Fermi surface~\cite{motrunich2005variational}.

\emph{Muon spin relaxation rate.}---Spin relaxation rates $\lambda$ provide further insight into the GS spin dynamics of the system, because dynamic local magnetic field is induced by the effective spin-1/2 moment of Yb$^{3+}$ on the triangular lattice. The ZF-$\mu^+$ spin relaxation rate can be expressed as~\cite{keren2004dynamic,keren2001probing,bono2004mu}:
\begin{equation}
\lambda^{\rm ZF} = \gamma_{\mu}^2\int_0^{\infty}\langle \textbf{B}_{\rm loc}^{\perp}(t)\cdot\textbf{B}_{\rm loc}^{\perp}(0)\rangle dt\propto S^{\perp}_{\omega\rightarrow 0},
\label{relaxation rate3}
\end{equation}
where $S^{\perp}_{\omega\rightarrow 0}=\int_0^{\infty}\langle\textbf{s}_{i}^{\perp}(t)\cdot\textbf{s}_{i}^{\perp}(0)\rangle dt$, is the static spin structure factor with the sum over \textbf{q}. It is fully compatible with the static susceptibility at $\mathbf q=0$~\cite{avella2013strongly}.

At high temperatures ($T>4$\,K), the $\mu^+$ spin relaxations (Fig.~\ref{fig2}) can be well fitted by Eq.~\eqref{function1} with a stretching exponent $\beta\sim 1$ (see the inset of Fig.~\ref{fig4}), indicating a concentrated spin system (spin-1/2 triangular lattice of Yb$^{3+}$) with very fast spin fluctuations~\cite{uemura1985muon,uemura1994spin}. In this case, a Gaussian distribution of local magnetic fields is expected with the width $\Delta\sim\gamma_{\mu}\sqrt{\langle B_{\rm loc}^2\rangle}$ and $\langle B_{\rm loc}\rangle\sim 0$. In the high-$T$ (mean-field) limit, the Yb$^{3+}$ spin fluctuation rate can be estimated as $\upsilon=\sqrt{z}J_0s/h\sim 4\times10^{10}$\,Hz~\cite{uemura1994spin}, where $z=6$ is the coordination number. 

Above 4\,K, experimental $\mu^+$ spin relaxation rates, both $\lambda_{\parallel}^{\rm ZF}$ and $\lambda_{\perp}^{\rm ZF}$ with the incident $\mu^+$ polarization parallel and perpendicular to the \emph{c}-axis, reach temperature-independent values of 0.185(2) and 0.211(1) $\mu$s$^{-1}$, respectively (see Fig.~\ref{fig5}). Using Eq.~\eqref{relaxation rate1} with $H=0$:
\begin{equation}
\lambda(T > 4\,{\rm K},H) = \frac{2\Delta^2\nu}{\nu^2+(\mu_0 H\gamma_{\mu})^2},
\label{relaxation rate1}
\end{equation}
we can estimate the distribution width of the local magnetic fields, $\Delta\sim 6\times 10^{7}$\,Hz $\ll\upsilon$, confirming the fast spin fluctuation limit~\cite{uemura1994spin,bono2004mu}.

As the temperature goes down, all of the $\mu^+$ spin relaxation rates gradually increase by about 50\,\% compared to the high-$T$ limit and get saturated below 0.4\,K (see Fig.~\ref{fig5}). These observations indicate the slowing down of spin fluctuations and the development of spin correlations. At the lowest temperature of our measurement, $T\sim 0.07$\,K, $\lambda_\parallel^{\rm LF}(H)$ does not follow the Eq.~\eqref{relaxation rate1} (see Fig.~\ref{fig4}a), suggesting that the spin dynamic autocorrelation function $S(t)$ should take a general form, $S(t)\sim (\tau/t)^x\exp(-\upsilon t)$, instead of the simple exponential form ($x=0$)~\cite{keren2001probing,keren2004dynamic,kermarrec2011spin}, where $\tau$ and 1/$\upsilon$ are the early and late time cutoffs, respectively, and $x$ can be defined as a critical exponent~\cite{hohenberg1977theory}. 

A more general expression for the $\mu^+$ spin relaxation rate can be obtained from both semi-classical~\cite{keren2001probing} and full quantum~\cite{mcmullen1978positive} treatments,
\begin{equation}
\lambda(H) = 2\Delta^2\tau^x\int^{\infty}_0t^{-x}\exp(-\upsilon t)\cos(2\pi\mu_0\gamma_{\mu}Ht)dt.
\label{relaxation rate2}
\end{equation}
For a spin system at $T\gg J_0$, the $\mu^+$ relaxation rates take the simple form of Eq.~\eqref{relaxation rate1}. On the other hand, at 0.07\,K $\lambda_\parallel^{\rm LF}(H)$ is much better described by Eq.~\eqref{relaxation rate2} with $x=0.66(5)$ (see Fig.~\ref{fig4}a), suggesting the onset of long-time spin correlations. At $T<0.4$\,K, the stretching exponent $\beta$ approaches a constant value of $\sim$ 0.6, indicating a distribution of the $\mu^+$ spin relaxation rates.

\emph{Phase diagram and conclusions.}---Temperature evolution of both bulk static susceptibilities and $\mu^+$ spin relaxation rates reveals several distinct regimes of YbMgGaO$_4$, as shown on Fig.~\ref{fig5}. At $T>4$\,K, the susceptibilities follow the Curie-Weiss law~\cite{li2015rare}, and the $\mu^+$ spin relaxation rates are constants, as typical for the paramagnetic regime~\cite{uemura1994spin,keren2004dynamic,clark2013gapless,faak2012kapellasite,gomilsek2016mu,gomilvsek2016instabilities}. The spins are still short-time correlated with $S(t)\sim\exp(-\upsilon t)$. This high-temperature phase is compatible with the classical paramagnet. As the temperature goes down (0.4\,K$<T<$4\,K), the spin susceptibilities show an unconventional critical behavior $\chi\sim T^{-1/3}$, and the $\mu^+$ spin relaxation rates increase by 50\% following the uniform spin susceptibilities (Fig.~\ref{fig5}). 

As the temperature is decreased even further ($T<0.4$\,K), the $\mu^+$ spin relaxation rates approach constant values again (Fig.~\ref{fig5}). This behavior appears to be generic for frustrated magnets~\cite{uemura1994spin,keren2004dynamic,clark2013gapless,faak2012kapellasite,gomilsek2016mu,gomilvsek2016instabilities}.
The temperature-independent $\mu^+$ spin relaxation rates suggest a constant spin correlation $S_{\omega\rightarrow 0}$ with the sum over \textbf{q}, which is consistent with the triangular U(1) QSL GS~\cite{motrunich2005variational,lee2005u,motrunich2006orbital}. Furthermore, the Yb$^{3+}$ spins on the triangular lattice become long-time correlated, $S(t)\sim (\tau/t)^x\exp(-\upsilon t)$, with $x=0.66(5)$.

In conclusion, we performed a comprehensive study of spin dynamics in the frustrated antiferromagnet YbMgGaO$_{4}$ by $\mu$SR measurements using the high-quality single crystals along both \emph{c}-axis and \emph{ab}-plane. No static uniform or random field is detected ($\langle B_{\rm loc}\rangle<0.9$\,Oe), indicating the absence of spin freezing ($\langle\mathbf{s}_{j}\rangle\sim 0$) down to at least 0.048\,K. Below 0.4\,K, both local susceptibility (Knight shift) and spin correlation function reach constant values. Long-time Yb$^{3+}$ spin correlations are developed in this temperature range. Combined with the heat-capacity data~\cite{li2015gapless}, our observations provide compelling evidence for the formation of gapless U(1) QSL GS in the triangular antiferromagnet YbMgGaO$_{4}$.

The concept of the U(1) QSL has been applied to a gamut of systems, from spin-ice pyrochlores~\cite{hermele2004,savary2012} to high-temperature superconductors~\cite{lee2008}, but experimental observation of this state remains elusive. Spin liquids in the organic charge-transfer salts with the frustrated triangular geometry deviate from the anticipated U(1) behavior~\cite{yamashita2008thermodynamic,yamashita2011gapless,motrunich2005variational}. In contrast, YbMgGaO$_4$ displays several phenomenological signatures of this state: i) the absence of long-range magnetic order and spin freezing; ii) the constant magnetic susceptibility at low temperatures; iii) the $C_v\simeq T^{2/3}$ power-law behavior of the specific heat~\cite{li2015gapless}. Our findings pave the way for observing further emergent properties of the triangular U(1) QSL, including the $\kappa\simeq T^{1/3}$ behavior of thermal conductivity and violation of the Wiedemann-Franz law~\cite{lee2005u,nave2007}, the power-law optical conductivity inside the Mott gap~\cite{ng2007}, and surface plasmons driven by spinons~\cite{ma2015}. Exact origin of the U(1) QSL GS of YbMgGaO$_4$ is also of interest. In contrast to the organic charge-transfer salts, where long-range magnetic order is destabilized by ring exchange, spin-orbit coupling and the ensuing magnetic anisotropy are the most likely effects that trigger strong frustration in this material~\cite{li2015}.

\emph{Acknowledgment}---We thank Gang Chen for confirming that the temperature-independent S$_{\omega\rightarrow 0}$ at low temperatures is consistent with the U(1) QSL GS. We thank Yipeng Cai and Adrian Hillier for helpful discussion. This work was supported by the NSF of China and the Ministry of Science and Technology of China (973 projects: 2012CB921701).
Y.S.L. was supported by the start-up funds of Renmin University of China. The work in Augsburg was supported by German Federal Ministry for Education and Research through the Sofja Kovalevskaya Award of Alexander von Humboldt Foundation. Q.M.Z. was supported by the Fundamental Research Funds for the Central Universities, and by the Research Funds of Renmin University of China.

\bibliography{MUSR_YbMgGaO4}

\begin{thebibliography}{59}%
\makeatletter
\providecommand \@ifxundefined [1]{%
 \@ifx{#1\undefined}
}%
\providecommand \@ifnum [1]{%
 \ifnum #1\expandafter \@firstoftwo
 \else \expandafter \@secondoftwo
 \fi
}%
\providecommand \@ifx [1]{%
 \ifx #1\expandafter \@firstoftwo
 \else \expandafter \@secondoftwo
 \fi
}%
\providecommand \natexlab [1]{#1}%
\providecommand \enquote  [1]{``#1''}%
\providecommand \bibnamefont  [1]{#1}%
\providecommand \bibfnamefont [1]{#1}%
\providecommand \citenamefont [1]{#1}%
\providecommand \href@noop [0]{\@secondoftwo}%
\providecommand \href [0]{\begingroup \@sanitize@url \@href}%
\providecommand \@href[1]{\@@startlink{#1}\@@href}%
\providecommand \@@href[1]{\endgroup#1\@@endlink}%
\providecommand \@sanitize@url [0]{\catcode `\\12\catcode `\$12\catcode
  `\&12\catcode `\#12\catcode `\^12\catcode `\_12\catcode `\%12\relax}%
\providecommand \@@startlink[1]{}%
\providecommand \@@endlink[0]{}%
\providecommand \url  [0]{\begingroup\@sanitize@url \@url }%
\providecommand \@url [1]{\endgroup\@href {#1}{\urlprefix }}%
\providecommand \urlprefix  [0]{URL }%
\providecommand \Eprint [0]{\href }%
\providecommand \doibase [0]{http://dx.doi.org/}%
\providecommand \selectlanguage [0]{\@gobble}%
\providecommand \bibinfo  [0]{\@secondoftwo}%
\providecommand \bibfield  [0]{\@secondoftwo}%
\providecommand \translation [1]{[#1]}%
\providecommand \BibitemOpen [0]{}%
\providecommand \bibitemStop [0]{}%
\providecommand \bibitemNoStop [0]{.\EOS\space}%
\providecommand \EOS [0]{\spacefactor3000\relax}%
\providecommand \BibitemShut  [1]{\csname bibitem#1\endcsname}%
\let\auto@bib@innerbib\@empty
\bibitem [{\citenamefont {Balents}(2010)}]{balents2010spin}%
  \BibitemOpen
  \bibfield  {author} {\bibinfo {author} {\bibfnamefont {L.}~\bibnamefont
  {Balents}},\ }\bibfield  {title} {\enquote {\bibinfo {title} {Spin liquids in
  frustrated magnets},}\ }\href@noop {} {\bibfield  {journal} {\bibinfo
  {journal} {Nature}\ }\textbf {\bibinfo {volume} {464}},\ \bibinfo {pages}
  {199--208} (\bibinfo {year} {2010})}\BibitemShut {NoStop}%
\bibitem [{\citenamefont {Lee}(2008{\natexlab{a}})}]{lee2008end}%
  \BibitemOpen
  \bibfield  {author} {\bibinfo {author} {\bibfnamefont {P.~A.}\ \bibnamefont
  {Lee}},\ }\bibfield  {title} {\enquote {\bibinfo {title} {An end to the
  drought of quantum spin liquids},}\ }\href@noop {} {\bibfield  {journal}
  {\bibinfo  {journal} {Science}\ }\textbf {\bibinfo {volume} {321}},\ \bibinfo
  {pages} {1306--1307} (\bibinfo {year} {2008}{\natexlab{a}})}\BibitemShut
  {NoStop}%
\bibitem [{\citenamefont {Wen}(2004)}]{wen2004quantum}%
  \BibitemOpen
  \bibfield  {author} {\bibinfo {author} {\bibfnamefont {X.-G.}\ \bibnamefont
  {Wen}},\ }\href@noop {} {\emph {\bibinfo {title} {Quantum field theory of
  many-body systems: from the origin of sound to an origin of light and
  electrons}}}\ (\bibinfo  {publisher} {Oxford University Press on Demand},\
  \bibinfo {year} {2004})\BibitemShut {NoStop}%
\bibitem [{\citenamefont {Anderson}(1973)}]{anderson1973resonating}%
  \BibitemOpen
  \bibfield  {author} {\bibinfo {author} {\bibfnamefont {P.~W.}\ \bibnamefont
  {Anderson}},\ }\bibfield  {title} {\enquote {\bibinfo {title} {Resonating
  valence bonds: A new kind of insulator?}}\ }\href@noop {} {\bibfield
  {journal} {\bibinfo  {journal} {Mater. Res. Bull.}\ }\textbf {\bibinfo
  {volume} {8}},\ \bibinfo {pages} {153--160} (\bibinfo {year}
  {1973})}\BibitemShut {NoStop}%
\bibitem [{\citenamefont {Anderson}(1987)}]{anderson1987resonating}%
  \BibitemOpen
  \bibfield  {author} {\bibinfo {author} {\bibfnamefont {P.~W.}\ \bibnamefont
  {Anderson}},\ }\bibfield  {title} {\enquote {\bibinfo {title} {The resonating
  valence bond state in {La$_2$CuO$_4$} and superconductivity},}\ }\href@noop
  {} {\bibfield  {journal} {\bibinfo  {journal} {Science}\ }\textbf {\bibinfo
  {volume} {235}},\ \bibinfo {pages} {1196--1198} (\bibinfo {year}
  {1987})}\BibitemShut {NoStop}%
\bibitem [{\citenamefont {Baskaran}\ \emph {et~al.}(1987)\citenamefont
  {Baskaran}, \citenamefont {Zou},\ and\ \citenamefont
  {Anderson}}]{baskaran1987resonating}%
  \BibitemOpen
  \bibfield  {author} {\bibinfo {author} {\bibfnamefont {G.}~\bibnamefont
  {Baskaran}}, \bibinfo {author} {\bibfnamefont {Z.}~\bibnamefont {Zou}}, \
  and\ \bibinfo {author} {\bibfnamefont {P.W.}\ \bibnamefont {Anderson}},\
  }\bibfield  {title} {\enquote {\bibinfo {title} {The resonating valence bond
  state and high-$t_c$ superconductivity: A mean field theory},}\ }\href@noop
  {} {\bibfield  {journal} {\bibinfo  {journal} {Solid State Comm.}\ }\textbf
  {\bibinfo {volume} {63}},\ \bibinfo {pages} {973--976} (\bibinfo {year}
  {1987})}\BibitemShut {NoStop}%
\bibitem [{\citenamefont {Singh}\ and\ \citenamefont
  {Huse}(1992)}]{singh1992three}%
  \BibitemOpen
  \bibfield  {author} {\bibinfo {author} {\bibfnamefont {R.R.P.}\ \bibnamefont
  {Singh}}\ and\ \bibinfo {author} {\bibfnamefont {D.~A.}\ \bibnamefont
  {Huse}},\ }\bibfield  {title} {\enquote {\bibinfo {title} {Three-sublattice
  order in triangular-and kagom{\'e}-lattice spin-half antiferromagnets},}\
  }\href@noop {} {\bibfield  {journal} {\bibinfo  {journal} {Phys. Rev. Lett.}\
  }\textbf {\bibinfo {volume} {68}},\ \bibinfo {pages} {1766} (\bibinfo {year}
  {1992})}\BibitemShut {NoStop}%
\bibitem [{\citenamefont {Sindzingre}\ \emph {et~al.}(1994)\citenamefont
  {Sindzingre}, \citenamefont {Lecheminant},\ and\ \citenamefont
  {Lhuillier}}]{sindzingre1994investigation}%
  \BibitemOpen
  \bibfield  {author} {\bibinfo {author} {\bibfnamefont {P.}~\bibnamefont
  {Sindzingre}}, \bibinfo {author} {\bibfnamefont {P.}~\bibnamefont
  {Lecheminant}}, \ and\ \bibinfo {author} {\bibfnamefont {C.}~\bibnamefont
  {Lhuillier}},\ }\bibfield  {title} {\enquote {\bibinfo {title} {Investigation
  of different classes of variational functions for the triangular and kagom\'e
  spin-1/2 {H}eisenberg antiferromagnets},}\ }\href@noop {} {\bibfield
  {journal} {\bibinfo  {journal} {Phys. Rev. B}\ }\textbf {\bibinfo {volume}
  {50}},\ \bibinfo {pages} {3108} (\bibinfo {year} {1994})}\BibitemShut
  {NoStop}%
\bibitem [{\citenamefont {Bernu}\ \emph {et~al.}(1994)\citenamefont {Bernu},
  \citenamefont {Lecheminant}, \citenamefont {Lhuillier},\ and\ \citenamefont
  {Pierre}}]{bernu1994exact}%
  \BibitemOpen
  \bibfield  {author} {\bibinfo {author} {\bibfnamefont {B.}~\bibnamefont
  {Bernu}}, \bibinfo {author} {\bibfnamefont {P.}~\bibnamefont {Lecheminant}},
  \bibinfo {author} {\bibfnamefont {C.}~\bibnamefont {Lhuillier}}, \ and\
  \bibinfo {author} {\bibfnamefont {L.}~\bibnamefont {Pierre}},\ }\bibfield
  {title} {\enquote {\bibinfo {title} {Exact spectra, spin susceptibilities,
  and order parameter of the quantum {H}eisenberg antiferromagnet on the
  triangular lattice},}\ }\href@noop {} {\bibfield  {journal} {\bibinfo
  {journal} {Phys. Rev. B}\ }\textbf {\bibinfo {volume} {50}},\ \bibinfo
  {pages} {10048} (\bibinfo {year} {1994})}\BibitemShut {NoStop}%
\bibitem [{\citenamefont {Capriotti}\ \emph {et~al.}(1999)\citenamefont
  {Capriotti}, \citenamefont {Trumper},\ and\ \citenamefont
  {Sorella}}]{capriotti1999long}%
  \BibitemOpen
  \bibfield  {author} {\bibinfo {author} {\bibfnamefont {L.}~\bibnamefont
  {Capriotti}}, \bibinfo {author} {\bibfnamefont {A.~E.}\ \bibnamefont
  {Trumper}}, \ and\ \bibinfo {author} {\bibfnamefont {S.}~\bibnamefont
  {Sorella}},\ }\bibfield  {title} {\enquote {\bibinfo {title} {Long-range
  {N}\'eel order in the triangular {H}eisenberg model},}\ }\href@noop {}
  {\bibfield  {journal} {\bibinfo  {journal} {Phys. Rev. Lett.}\ }\textbf
  {\bibinfo {volume} {82}},\ \bibinfo {pages} {3899} (\bibinfo {year}
  {1999})}\BibitemShut {NoStop}%
\bibitem [{\citenamefont {Weihong}\ \emph {et~al.}(1999)\citenamefont
  {Weihong}, \citenamefont {McKenzie},\ and\ \citenamefont
  {Singh}}]{weihong1999phase}%
  \BibitemOpen
  \bibfield  {author} {\bibinfo {author} {\bibfnamefont {Z.}~\bibnamefont
  {Weihong}}, \bibinfo {author} {\bibfnamefont {R.~H.}\ \bibnamefont
  {McKenzie}}, \ and\ \bibinfo {author} {\bibfnamefont {R.R.P.}\ \bibnamefont
  {Singh}},\ }\bibfield  {title} {\enquote {\bibinfo {title} {Phase diagram for
  a class of spin-$\frac12$ {H}eisenberg models interpolating between the
  square-lattice, the triangular-lattice, and the linear-chain limits},}\
  }\href@noop {} {\bibfield  {journal} {\bibinfo  {journal} {Phys. Rev. B}\
  }\textbf {\bibinfo {volume} {59}},\ \bibinfo {pages} {14367} (\bibinfo {year}
  {1999})}\BibitemShut {NoStop}%
\bibitem [{\citenamefont {Kaneko}\ \emph {et~al.}(2014)\citenamefont {Kaneko},
  \citenamefont {Morita},\ and\ \citenamefont {Imada}}]{kaneko2014gapless}%
  \BibitemOpen
  \bibfield  {author} {\bibinfo {author} {\bibfnamefont {R.}~\bibnamefont
  {Kaneko}}, \bibinfo {author} {\bibfnamefont {S.}~\bibnamefont {Morita}}, \
  and\ \bibinfo {author} {\bibfnamefont {M.}~\bibnamefont {Imada}},\ }\bibfield
   {title} {\enquote {\bibinfo {title} {Gapless spin-liquid phase in an
  extended spin-$\frac12$ triangular {H}eisenberg model},}\ }\href@noop {}
  {\bibfield  {journal} {\bibinfo  {journal} {J. Phys. Soc. Jpn.}\ }\textbf
  {\bibinfo {volume} {83}},\ \bibinfo {pages} {093707} (\bibinfo {year}
  {2014})}\BibitemShut {NoStop}%
\bibitem [{\citenamefont {Li}\ \emph {et~al.}(2015{\natexlab{a}})\citenamefont
  {Li}, \citenamefont {Bishop},\ and\ \citenamefont
  {Campbell}}]{li2015quasiclassical}%
  \BibitemOpen
  \bibfield  {author} {\bibinfo {author} {\bibfnamefont {P.H.Y.}\ \bibnamefont
  {Li}}, \bibinfo {author} {\bibfnamefont {R.~F.}\ \bibnamefont {Bishop}}, \
  and\ \bibinfo {author} {\bibfnamefont {C.~E.}\ \bibnamefont {Campbell}},\
  }\bibfield  {title} {\enquote {\bibinfo {title} {Quasiclassical magnetic
  order and its loss in a spin-$\frac12$ {H}eisenberg antiferromagnet on a
  triangular lattice with competing bonds},}\ }\href@noop {} {\bibfield
  {journal} {\bibinfo  {journal} {Phys. Rev. B}\ }\textbf {\bibinfo {volume}
  {91}},\ \bibinfo {pages} {014426} (\bibinfo {year}
  {2015}{\natexlab{a}})}\BibitemShut {NoStop}%
\bibitem [{\citenamefont {Trumper}(1999)}]{trumper1999spin}%
  \BibitemOpen
  \bibfield  {author} {\bibinfo {author} {\bibfnamefont {A.~E.}\ \bibnamefont
  {Trumper}},\ }\bibfield  {title} {\enquote {\bibinfo {title} {Spin-wave
  analysis to the spatially anisotropic {H}eisenberg antiferromagnet on a
  triangular lattice},}\ }\href@noop {} {\bibfield  {journal} {\bibinfo
  {journal} {Phys. Rev. B}\ }\textbf {\bibinfo {volume} {60}},\ \bibinfo
  {pages} {2987} (\bibinfo {year} {1999})}\BibitemShut {NoStop}%
\bibitem [{\citenamefont {Yunoki}\ and\ \citenamefont
  {Sorella}(2006)}]{yunoki2006two}%
  \BibitemOpen
  \bibfield  {author} {\bibinfo {author} {\bibfnamefont {S.}~\bibnamefont
  {Yunoki}}\ and\ \bibinfo {author} {\bibfnamefont {S.}~\bibnamefont
  {Sorella}},\ }\bibfield  {title} {\enquote {\bibinfo {title} {Two spin liquid
  phases in the spatially anisotropic triangular {H}eisenberg model},}\
  }\href@noop {} {\bibfield  {journal} {\bibinfo  {journal} {Phys. Rev. B}\
  }\textbf {\bibinfo {volume} {74}},\ \bibinfo {pages} {014408} (\bibinfo
  {year} {2006})}\BibitemShut {NoStop}%
\bibitem [{\citenamefont {Ohashi}\ \emph {et~al.}(2008)\citenamefont {Ohashi},
  \citenamefont {Momoi}, \citenamefont {Tsunetsugu},\ and\ \citenamefont
  {Kawakami}}]{ohashi2008finite}%
  \BibitemOpen
  \bibfield  {author} {\bibinfo {author} {\bibfnamefont {T.}~\bibnamefont
  {Ohashi}}, \bibinfo {author} {\bibfnamefont {T.}~\bibnamefont {Momoi}},
  \bibinfo {author} {\bibfnamefont {H.}~\bibnamefont {Tsunetsugu}}, \ and\
  \bibinfo {author} {\bibfnamefont {N.}~\bibnamefont {Kawakami}},\ }\bibfield
  {title} {\enquote {\bibinfo {title} {Finite temperature {M}ott transition in
  {H}ubbard model on anisotropic triangular lattice},}\ }\href@noop {}
  {\bibfield  {journal} {\bibinfo  {journal} {Phys. Rev. Lett.}\ }\textbf
  {\bibinfo {volume} {100}},\ \bibinfo {pages} {076402} (\bibinfo {year}
  {2008})}\BibitemShut {NoStop}%
\bibitem [{\citenamefont {Watanabe}\ \emph {et~al.}(2014)\citenamefont
  {Watanabe}, \citenamefont {Kawamura}, \citenamefont {Nakano},\ and\
  \citenamefont {Sakai}}]{watanabe2014quantum}%
  \BibitemOpen
  \bibfield  {author} {\bibinfo {author} {\bibfnamefont {K.}~\bibnamefont
  {Watanabe}}, \bibinfo {author} {\bibfnamefont {H.}~\bibnamefont {Kawamura}},
  \bibinfo {author} {\bibfnamefont {H.}~\bibnamefont {Nakano}}, \ and\ \bibinfo
  {author} {\bibfnamefont {T.}~\bibnamefont {Sakai}},\ }\bibfield  {title}
  {\enquote {\bibinfo {title} {Quantum spin-liquid behavior in the spin-1/2
  random {H}eisenberg antiferromagnet on the triangular lattice},}\ }\href@noop
  {} {\bibfield  {journal} {\bibinfo  {journal} {J. Phys. Soc. Jpn.}\ }\textbf
  {\bibinfo {volume} {83}},\ \bibinfo {pages} {034714} (\bibinfo {year}
  {2014})}\BibitemShut {NoStop}%
\bibitem [{\citenamefont {Motrunich}(2005)}]{motrunich2005variational}%
  \BibitemOpen
  \bibfield  {author} {\bibinfo {author} {\bibfnamefont {O.~I}\ \bibnamefont
  {Motrunich}},\ }\bibfield  {title} {\enquote {\bibinfo {title} {Variational
  study of triangular lattice spin-$\frac12$ model with ring exchanges and spin
  liquid state in {$\kappa$-(ET)$_2$Cu$_2$(CN)$_3$}},}\ }\href@noop {}
  {\bibfield  {journal} {\bibinfo  {journal} {Phys. Rev. B}\ }\textbf {\bibinfo
  {volume} {72}},\ \bibinfo {pages} {045105} (\bibinfo {year}
  {2005})}\BibitemShut {NoStop}%
\bibitem [{\citenamefont {Lee}\ and\ \citenamefont {Lee}(2005)}]{lee2005u}%
  \BibitemOpen
  \bibfield  {author} {\bibinfo {author} {\bibfnamefont {S.-S.}\ \bibnamefont
  {Lee}}\ and\ \bibinfo {author} {\bibfnamefont {P.A.}\ \bibnamefont {Lee}},\
  }\bibfield  {title} {\enquote {\bibinfo {title} {U(1) gauge theory of the
  {H}ubbard model: Spin liquid states and possible application to
  {$\kappa$-(BEDT-TTF)$_2$Cu$_2$(CN)$_3$}},}\ }\href@noop {} {\bibfield
  {journal} {\bibinfo  {journal} {Phys. Rev. Lett.}\ }\textbf {\bibinfo
  {volume} {95}},\ \bibinfo {pages} {036403} (\bibinfo {year}
  {2005})}\BibitemShut {NoStop}%
\bibitem [{\citenamefont {Motrunich}(2006)}]{motrunich2006orbital}%
  \BibitemOpen
  \bibfield  {author} {\bibinfo {author} {\bibfnamefont {O.~I.}\ \bibnamefont
  {Motrunich}},\ }\bibfield  {title} {\enquote {\bibinfo {title} {Orbital
  magnetic field effects in spin liquid with spinon {F}ermi sea: Possible
  application to {$\kappa$-(ET)$_2$Cu$_2$(CN)$_3$}},}\ }\href@noop {}
  {\bibfield  {journal} {\bibinfo  {journal} {Phys. Rev. B}\ }\textbf {\bibinfo
  {volume} {73}},\ \bibinfo {pages} {155115} (\bibinfo {year}
  {2006})}\BibitemShut {NoStop}%
\bibitem [{\citenamefont {Yamashita}\ \emph {et~al.}(2008)\citenamefont
  {Yamashita}, \citenamefont {Nakazawa}, \citenamefont {Oguni}, \citenamefont
  {Oshima}, \citenamefont {Nojiri}, \citenamefont {Shimizu}, \citenamefont
  {Miyagawa},\ and\ \citenamefont {Kanoda}}]{yamashita2008thermodynamic}%
  \BibitemOpen
  \bibfield  {author} {\bibinfo {author} {\bibfnamefont {S.}~\bibnamefont
  {Yamashita}}, \bibinfo {author} {\bibfnamefont {Y.}~\bibnamefont {Nakazawa}},
  \bibinfo {author} {\bibfnamefont {M.}~\bibnamefont {Oguni}}, \bibinfo
  {author} {\bibfnamefont {Y.}~\bibnamefont {Oshima}}, \bibinfo {author}
  {\bibfnamefont {H.}~\bibnamefont {Nojiri}}, \bibinfo {author} {\bibfnamefont
  {Y.}~\bibnamefont {Shimizu}}, \bibinfo {author} {\bibfnamefont
  {K.}~\bibnamefont {Miyagawa}}, \ and\ \bibinfo {author} {\bibfnamefont
  {K.}~\bibnamefont {Kanoda}},\ }\bibfield  {title} {\enquote {\bibinfo {title}
  {Thermodynamic properties of a spin-1/2 spin-liquid state in a $\kappa$-type
  organic salt},}\ }\href@noop {} {\bibfield  {journal} {\bibinfo  {journal}
  {Nature Phys.}\ }\textbf {\bibinfo {volume} {4}},\ \bibinfo {pages}
  {459--462} (\bibinfo {year} {2008})}\BibitemShut {NoStop}%
\bibitem [{\citenamefont {Yamashita}\ \emph {et~al.}(2011)\citenamefont
  {Yamashita}, \citenamefont {Yamamoto}, \citenamefont {Nakazawa},
  \citenamefont {Tamura},\ and\ \citenamefont {Kato}}]{yamashita2011gapless}%
  \BibitemOpen
  \bibfield  {author} {\bibinfo {author} {\bibfnamefont {S.}~\bibnamefont
  {Yamashita}}, \bibinfo {author} {\bibfnamefont {T.}~\bibnamefont {Yamamoto}},
  \bibinfo {author} {\bibfnamefont {Y.}~\bibnamefont {Nakazawa}}, \bibinfo
  {author} {\bibfnamefont {M.}~\bibnamefont {Tamura}}, \ and\ \bibinfo {author}
  {\bibfnamefont {R.}~\bibnamefont {Kato}},\ }\bibfield  {title} {\enquote
  {\bibinfo {title} {Gapless spin liquid of an organic triangular compound
  evidenced by thermodynamic measurements},}\ }\href@noop {} {\bibfield
  {journal} {\bibinfo  {journal} {Nature Comm.}\ }\textbf {\bibinfo {volume}
  {2}},\ \bibinfo {pages} {275} (\bibinfo {year} {2011})}\BibitemShut {NoStop}%
\bibitem [{\citenamefont {Li}\ \emph {et~al.}(2015{\natexlab{b}})\citenamefont
  {Li}, \citenamefont {Liao}, \citenamefont {Zhang}, \citenamefont {Li},
  \citenamefont {Jin}, \citenamefont {Ling}, \citenamefont {Zhang},
  \citenamefont {Zou}, \citenamefont {Pi}, \citenamefont {Yang} \emph
  {et~al.}}]{li2015gapless}%
  \BibitemOpen
  \bibfield  {author} {\bibinfo {author} {\bibfnamefont {Y.}~\bibnamefont
  {Li}}, \bibinfo {author} {\bibfnamefont {H.}~\bibnamefont {Liao}}, \bibinfo
  {author} {\bibfnamefont {Z.}~\bibnamefont {Zhang}}, \bibinfo {author}
  {\bibfnamefont {S.}~\bibnamefont {Li}}, \bibinfo {author} {\bibfnamefont
  {F.}~\bibnamefont {Jin}}, \bibinfo {author} {\bibfnamefont {L.}~\bibnamefont
  {Ling}}, \bibinfo {author} {\bibfnamefont {L.}~\bibnamefont {Zhang}},
  \bibinfo {author} {\bibfnamefont {Y.}~\bibnamefont {Zou}}, \bibinfo {author}
  {\bibfnamefont {L.}~\bibnamefont {Pi}}, \bibinfo {author} {\bibfnamefont
  {Z.}~\bibnamefont {Yang}},  \emph {et~al.},\ }\bibfield  {title} {\enquote
  {\bibinfo {title} {Gapless quantum spin liquid ground state in the
  two-dimensional spin-1/2 triangular antiferromagnet {YbMgGaO$_4$}},}\
  }\href@noop {} {\bibfield  {journal} {\bibinfo  {journal} {Scientific
  Reports}\ }\textbf {\bibinfo {volume} {5}} (\bibinfo {year}
  {2015}{\natexlab{b}})}\BibitemShut {NoStop}%
\bibitem [{\citenamefont {Li}\ \emph {et~al.}(2015{\natexlab{c}})\citenamefont
  {Li}, \citenamefont {Chen}, \citenamefont {Tong}, \citenamefont {Pi},
  \citenamefont {Liu}, \citenamefont {Yang}, \citenamefont {Wang},\ and\
  \citenamefont {Zhang}}]{li2015rare}%
  \BibitemOpen
  \bibfield  {author} {\bibinfo {author} {\bibfnamefont {Y.}~\bibnamefont
  {Li}}, \bibinfo {author} {\bibfnamefont {G.}~\bibnamefont {Chen}}, \bibinfo
  {author} {\bibfnamefont {W.}~\bibnamefont {Tong}}, \bibinfo {author}
  {\bibfnamefont {L.}~\bibnamefont {Pi}}, \bibinfo {author} {\bibfnamefont
  {J.}~\bibnamefont {Liu}}, \bibinfo {author} {\bibfnamefont {Z.}~\bibnamefont
  {Yang}}, \bibinfo {author} {\bibfnamefont {X.}~\bibnamefont {Wang}}, \ and\
  \bibinfo {author} {\bibfnamefont {Q.}~\bibnamefont {Zhang}},\ }\bibfield
  {title} {\enquote {\bibinfo {title} {Rare-earth triangular lattice spin
  liquid: a single-crystal study of {YbMgGaO$_4$}},}\ }\href@noop {} {\bibfield
   {journal} {\bibinfo  {journal} {Phys. Rev. Lett.}\ }\textbf {\bibinfo
  {volume} {115}},\ \bibinfo {pages} {167203} (\bibinfo {year}
  {2015}{\natexlab{c}})}\BibitemShut {NoStop}%
\bibitem [{\citenamefont {Lee}\ \emph {et~al.}(2007)\citenamefont {Lee},
  \citenamefont {Kikuchi}, \citenamefont {Qiu}, \citenamefont {Lake},
  \citenamefont {Huang}, \citenamefont {Habicht},\ and\ \citenamefont
  {Kiefer}}]{lee2007quantum}%
  \BibitemOpen
  \bibfield  {author} {\bibinfo {author} {\bibfnamefont {S.-H.}\ \bibnamefont
  {Lee}}, \bibinfo {author} {\bibfnamefont {H.}~\bibnamefont {Kikuchi}},
  \bibinfo {author} {\bibfnamefont {Y.}~\bibnamefont {Qiu}}, \bibinfo {author}
  {\bibfnamefont {B.}~\bibnamefont {Lake}}, \bibinfo {author} {\bibfnamefont
  {Q.}~\bibnamefont {Huang}}, \bibinfo {author} {\bibfnamefont
  {K.}~\bibnamefont {Habicht}}, \ and\ \bibinfo {author} {\bibfnamefont
  {K.}~\bibnamefont {Kiefer}},\ }\bibfield  {title} {\enquote {\bibinfo {title}
  {Quantum-spin-liquid states in the two-dimensional kagome antiferromagnets
  {Zn$_x$Cu$_{4-x}$(OD)$_6$Cl$_2$}},}\ }\href@noop {} {\bibfield  {journal}
  {\bibinfo  {journal} {Nature Mater.}\ }\textbf {\bibinfo {volume} {6}},\
  \bibinfo {pages} {853--857} (\bibinfo {year} {2007})}\BibitemShut {NoStop}%
\bibitem [{\citenamefont {Freedman}\ \emph {et~al.}(2010)\citenamefont
  {Freedman}, \citenamefont {Han}, \citenamefont {Prodi}, \citenamefont
  {M\"uller}, \citenamefont {Huang}, \citenamefont {Chen}, \citenamefont
  {Webb}, \citenamefont {Lee}, \citenamefont {McQueen},\ and\ \citenamefont
  {Nocera}}]{freedman2010site}%
  \BibitemOpen
  \bibfield  {author} {\bibinfo {author} {\bibfnamefont {D.~E}\ \bibnamefont
  {Freedman}}, \bibinfo {author} {\bibfnamefont {T.~H.}\ \bibnamefont {Han}},
  \bibinfo {author} {\bibfnamefont {A.}~\bibnamefont {Prodi}}, \bibinfo
  {author} {\bibfnamefont {P.}~\bibnamefont {M\"uller}}, \bibinfo {author}
  {\bibfnamefont {Q.-Z.}\ \bibnamefont {Huang}}, \bibinfo {author}
  {\bibfnamefont {Y.-S.}\ \bibnamefont {Chen}}, \bibinfo {author}
  {\bibfnamefont {S.~M.}\ \bibnamefont {Webb}}, \bibinfo {author}
  {\bibfnamefont {Y.S.}\ \bibnamefont {Lee}}, \bibinfo {author} {\bibfnamefont
  {T.M.}\ \bibnamefont {McQueen}}, \ and\ \bibinfo {author} {\bibfnamefont
  {D.G.}\ \bibnamefont {Nocera}},\ }\bibfield  {title} {\enquote {\bibinfo
  {title} {Site specific x-ray anomalous dispersion of the geometrically
  frustrated kagome magnet, herbertsmithite, {ZnCu$_3$(OH)$_6$Cl$_2$}},}\
  }\href@noop {} {\bibfield  {journal} {\bibinfo  {journal} {J. Amer. Chem.
  Soc.}\ }\textbf {\bibinfo {volume} {132}},\ \bibinfo {pages} {16185--16190}
  (\bibinfo {year} {2010})}\BibitemShut {NoStop}%
\bibitem [{\citenamefont {Li}\ and\ \citenamefont
  {Zhang}(2012)}]{li2012structure}%
  \BibitemOpen
  \bibfield  {author} {\bibinfo {author} {\bibfnamefont {Y.}~\bibnamefont
  {Li}}\ and\ \bibinfo {author} {\bibfnamefont {Q.}~\bibnamefont {Zhang}},\
  }\bibfield  {title} {\enquote {\bibinfo {title} {Structure and magnetism of
  $s=\frac12$ kagome antiferromagnets {NiCu$_3$(OH)$_6$Cl$_2$ and
  CoCu$_3$(OH)$_6$Cl$_2$}},}\ }\href@noop {} {\bibfield  {journal} {\bibinfo
  {journal} {J. Phys.: Condens. Matter}\ }\textbf {\bibinfo {volume} {25}},\
  \bibinfo {pages} {026003} (\bibinfo {year} {2012})}\BibitemShut {NoStop}%
\bibitem [{\citenamefont {Li}\ \emph {et~al.}(2013)\citenamefont {Li},
  \citenamefont {Fu}, \citenamefont {Wu},\ and\ \citenamefont
  {Zhang}}]{li2013transition}%
  \BibitemOpen
  \bibfield  {author} {\bibinfo {author} {\bibfnamefont {Y.}~\bibnamefont
  {Li}}, \bibinfo {author} {\bibfnamefont {J.}~\bibnamefont {Fu}}, \bibinfo
  {author} {\bibfnamefont {Z.}~\bibnamefont {Wu}}, \ and\ \bibinfo {author}
  {\bibfnamefont {Q.}~\bibnamefont {Zhang}},\ }\bibfield  {title} {\enquote
  {\bibinfo {title} {Transition-metal distribution in kagome antiferromagnet
  {CoCu$_3$(OH)$_6$Cl$_2$} revealed by resonant x-ray diffraction},}\
  }\href@noop {} {\bibfield  {journal} {\bibinfo  {journal} {Chem. Phys.
  Lett.}\ }\textbf {\bibinfo {volume} {570}},\ \bibinfo {pages} {37--41}
  (\bibinfo {year} {2013})}\BibitemShut {NoStop}%
\bibitem [{\citenamefont {Li}\ \emph {et~al.}(2014)\citenamefont {Li},
  \citenamefont {Pan}, \citenamefont {Li}, \citenamefont {Tong}, \citenamefont
  {Ling}, \citenamefont {Yang}, \citenamefont {Wang}, \citenamefont {Chen},
  \citenamefont {Wu},\ and\ \citenamefont {Zhang}}]{li2014gapless}%
  \BibitemOpen
  \bibfield  {author} {\bibinfo {author} {\bibfnamefont {Y.}~\bibnamefont
  {Li}}, \bibinfo {author} {\bibfnamefont {B.}~\bibnamefont {Pan}}, \bibinfo
  {author} {\bibfnamefont {S.}~\bibnamefont {Li}}, \bibinfo {author}
  {\bibfnamefont {W.}~\bibnamefont {Tong}}, \bibinfo {author} {\bibfnamefont
  {L.}~\bibnamefont {Ling}}, \bibinfo {author} {\bibfnamefont {Z.}~\bibnamefont
  {Yang}}, \bibinfo {author} {\bibfnamefont {J.}~\bibnamefont {Wang}}, \bibinfo
  {author} {\bibfnamefont {Z.}~\bibnamefont {Chen}}, \bibinfo {author}
  {\bibfnamefont {Z.}~\bibnamefont {Wu}}, \ and\ \bibinfo {author}
  {\bibfnamefont {Q.}~\bibnamefont {Zhang}},\ }\bibfield  {title} {\enquote
  {\bibinfo {title} {Gapless quantum spin liquid in the $s=1/2$ anisotropic
  kagome antiferromagnet {ZnCu$_3$(OH)$_6$SO$_4$}},}\ }\href@noop {} {\bibfield
   {journal} {\bibinfo  {journal} {New J. Phys.}\ }\textbf {\bibinfo {volume}
  {16}},\ \bibinfo {pages} {093011} (\bibinfo {year} {2014})}\BibitemShut
  {NoStop}%
\bibitem [{\citenamefont {Shimizu}\ \emph {et~al.}(2003)\citenamefont
  {Shimizu}, \citenamefont {Miyagawa}, \citenamefont {Kanoda}, \citenamefont
  {Maesato},\ and\ \citenamefont {Saito}}]{shimizu2003spin}%
  \BibitemOpen
  \bibfield  {author} {\bibinfo {author} {\bibfnamefont {Y.}~\bibnamefont
  {Shimizu}}, \bibinfo {author} {\bibfnamefont {K.}~\bibnamefont {Miyagawa}},
  \bibinfo {author} {\bibfnamefont {K.}~\bibnamefont {Kanoda}}, \bibinfo
  {author} {\bibfnamefont {M.}~\bibnamefont {Maesato}}, \ and\ \bibinfo
  {author} {\bibfnamefont {G.}~\bibnamefont {Saito}},\ }\bibfield  {title}
  {\enquote {\bibinfo {title} {Spin liquid state in an organic mott insulator
  with a triangular lattice},}\ }\href@noop {} {\bibfield  {journal} {\bibinfo
  {journal} {Phys. Rev. Lett.}\ }\textbf {\bibinfo {volume} {91}},\ \bibinfo
  {pages} {107001} (\bibinfo {year} {2003})}\BibitemShut {NoStop}%
\bibitem [{\citenamefont {Itou}\ \emph {et~al.}(2008)\citenamefont {Itou},
  \citenamefont {Oyamada}, \citenamefont {Maegawa}, \citenamefont {Tamura},\
  and\ \citenamefont {Kato}}]{itou2008quantum}%
  \BibitemOpen
  \bibfield  {author} {\bibinfo {author} {\bibfnamefont {T.}~\bibnamefont
  {Itou}}, \bibinfo {author} {\bibfnamefont {A.}~\bibnamefont {Oyamada}},
  \bibinfo {author} {\bibfnamefont {S.}~\bibnamefont {Maegawa}}, \bibinfo
  {author} {\bibfnamefont {M.}~\bibnamefont {Tamura}}, \ and\ \bibinfo {author}
  {\bibfnamefont {R.}~\bibnamefont {Kato}},\ }\bibfield  {title} {\enquote
  {\bibinfo {title} {Quantum spin liquid in the spin-1/2 triangular
  antiferromagnet {EtMe$_3$Sb[Pd(dmit)$_2$]$_2$}},}\ }\href@noop {} {\bibfield
  {journal} {\bibinfo  {journal} {Phys. Rev. B}\ }\textbf {\bibinfo {volume}
  {77}},\ \bibinfo {pages} {104413} (\bibinfo {year} {2008})}\BibitemShut
  {NoStop}%
\bibitem [{\citenamefont {Moriya}(1960)}]{moriya1960new}%
  \BibitemOpen
  \bibfield  {author} {\bibinfo {author} {\bibfnamefont {T.}~\bibnamefont
  {Moriya}},\ }\bibfield  {title} {\enquote {\bibinfo {title} {New mechanism of
  anisotropic superexchange interaction},}\ }\href@noop {} {\bibfield
  {journal} {\bibinfo  {journal} {Phys. Rev. Lett.}\ }\textbf {\bibinfo
  {volume} {4}},\ \bibinfo {pages} {228} (\bibinfo {year} {1960})}\BibitemShut
  {NoStop}%
\bibitem [{\citenamefont {Zorko}\ \emph {et~al.}(2008)\citenamefont {Zorko},
  \citenamefont {Nellutla}, \citenamefont {{Van Tol}}, \citenamefont {Brunel},
  \citenamefont {Bert}, \citenamefont {Duc}, \citenamefont {Trombe},
  \citenamefont {{De Vries}}, \citenamefont {Harrison},\ and\ \citenamefont
  {Mendels}}]{zorko2008dzyaloshinsky}%
  \BibitemOpen
  \bibfield  {author} {\bibinfo {author} {\bibfnamefont {A.}~\bibnamefont
  {Zorko}}, \bibinfo {author} {\bibfnamefont {S.}~\bibnamefont {Nellutla}},
  \bibinfo {author} {\bibfnamefont {J.}~\bibnamefont {{Van Tol}}}, \bibinfo
  {author} {\bibfnamefont {L.C.}\ \bibnamefont {Brunel}}, \bibinfo {author}
  {\bibfnamefont {F.}~\bibnamefont {Bert}}, \bibinfo {author} {\bibfnamefont
  {F.}~\bibnamefont {Duc}}, \bibinfo {author} {\bibfnamefont {J.-C.}\
  \bibnamefont {Trombe}}, \bibinfo {author} {\bibfnamefont {M.A.}\ \bibnamefont
  {{De Vries}}}, \bibinfo {author} {\bibfnamefont {A.}~\bibnamefont
  {Harrison}}, \ and\ \bibinfo {author} {\bibfnamefont {P.}~\bibnamefont
  {Mendels}},\ }\bibfield  {title} {\enquote {\bibinfo {title}
  {Dzyaloshinsky-{M}oriya anisotropy in the spin-1/2 kagome compound
  {ZnCu$_3$(OH)$_6$Cl$_2$}},}\ }\href@noop {} {\bibfield  {journal} {\bibinfo
  {journal} {Phys. Rev. Lett.}\ }\textbf {\bibinfo {volume} {101}},\ \bibinfo
  {pages} {026405} (\bibinfo {year} {2008})}\BibitemShut {NoStop}%
\bibitem [{\citenamefont {Bramwell}\ \emph {et~al.}(2009)\citenamefont
  {Bramwell}, \citenamefont {Giblin}, \citenamefont {Calder}, \citenamefont
  {Aldus}, \citenamefont {Prabhakaran},\ and\ \citenamefont
  {Fennell}}]{bramwell2009measurement}%
  \BibitemOpen
  \bibfield  {author} {\bibinfo {author} {\bibfnamefont {S.T.}\ \bibnamefont
  {Bramwell}}, \bibinfo {author} {\bibfnamefont {S.R.}\ \bibnamefont {Giblin}},
  \bibinfo {author} {\bibfnamefont {S.}~\bibnamefont {Calder}}, \bibinfo
  {author} {\bibfnamefont {R.}~\bibnamefont {Aldus}}, \bibinfo {author}
  {\bibfnamefont {D.}~\bibnamefont {Prabhakaran}}, \ and\ \bibinfo {author}
  {\bibfnamefont {T.}~\bibnamefont {Fennell}},\ }\bibfield  {title} {\enquote
  {\bibinfo {title} {Measurement of the charge and current of magnetic
  monopoles in spin ice},}\ }\href@noop {} {\bibfield  {journal} {\bibinfo
  {journal} {Nature}\ }\textbf {\bibinfo {volume} {461}},\ \bibinfo {pages}
  {956--959} (\bibinfo {year} {2009})}\BibitemShut {NoStop}%
\bibitem [{\citenamefont {Pratt}\ \emph {et~al.}(2011)\citenamefont {Pratt},
  \citenamefont {Baker}, \citenamefont {Blundell}, \citenamefont {Lancaster},
  \citenamefont {Ohira-Kawamura}, \citenamefont {Baines}, \citenamefont
  {Shimizu}, \citenamefont {Kanoda}, \citenamefont {Watanabe},\ and\
  \citenamefont {Saito}}]{pratt2011magnetic}%
  \BibitemOpen
  \bibfield  {author} {\bibinfo {author} {\bibfnamefont {F.L.}\ \bibnamefont
  {Pratt}}, \bibinfo {author} {\bibfnamefont {P.J.}\ \bibnamefont {Baker}},
  \bibinfo {author} {\bibfnamefont {S.J.}\ \bibnamefont {Blundell}}, \bibinfo
  {author} {\bibfnamefont {T.}~\bibnamefont {Lancaster}}, \bibinfo {author}
  {\bibfnamefont {S.}~\bibnamefont {Ohira-Kawamura}}, \bibinfo {author}
  {\bibfnamefont {C.}~\bibnamefont {Baines}}, \bibinfo {author} {\bibfnamefont
  {Y.}~\bibnamefont {Shimizu}}, \bibinfo {author} {\bibfnamefont
  {K.}~\bibnamefont {Kanoda}}, \bibinfo {author} {\bibfnamefont
  {I.}~\bibnamefont {Watanabe}}, \ and\ \bibinfo {author} {\bibfnamefont
  {G.}~\bibnamefont {Saito}},\ }\bibfield  {title} {\enquote {\bibinfo {title}
  {Magnetic and non-magnetic phases of a quantum spin liquid},}\ }\href@noop {}
  {\bibfield  {journal} {\bibinfo  {journal} {Nature}\ }\textbf {\bibinfo
  {volume} {471}},\ \bibinfo {pages} {612--616} (\bibinfo {year}
  {2011})}\BibitemShut {NoStop}%
\bibitem [{sup()}]{supple}%
  \BibitemOpen
  \href@noop {} {\emph {\bibinfo {title} {See Supplemental Material for
  detailed information about experimental procedures}}}\BibitemShut {NoStop}%
\bibitem [{\citenamefont {Hillier}\ \emph {et~al.}(2005)\citenamefont
  {Hillier}, \citenamefont {King}, \citenamefont {Cottrell},\ and\
  \citenamefont {Lord}}]{hillier2005musr}%
  \BibitemOpen
  \bibfield  {author} {\bibinfo {author} {\bibfnamefont {A.D.}\ \bibnamefont
  {Hillier}}, \bibinfo {author} {\bibfnamefont {P.J.C.}\ \bibnamefont {King}},
  \bibinfo {author} {\bibfnamefont {S.P.}\ \bibnamefont {Cottrell}}, \ and\
  \bibinfo {author} {\bibfnamefont {J.S.}\ \bibnamefont {Lord}},\ }\bibfield
  {title} {\enquote {\bibinfo {title} {The musr user guide},}\ }\href@noop {}
  {\bibfield  {journal} {\bibinfo  {journal} {ISIS Facility, STFC, Rutherford
  Appleton Laboratory, UK}\ } (\bibinfo {year} {2005})}\BibitemShut {NoStop}%
\bibitem [{\citenamefont {Keren}\ \emph {et~al.}(2004)\citenamefont {Keren},
  \citenamefont {Gardner}, \citenamefont {Ehlers}, \citenamefont {Fukaya},
  \citenamefont {Segal},\ and\ \citenamefont {Uemura}}]{keren2004dynamic}%
  \BibitemOpen
  \bibfield  {author} {\bibinfo {author} {\bibfnamefont {A.}~\bibnamefont
  {Keren}}, \bibinfo {author} {\bibfnamefont {J.S.}\ \bibnamefont {Gardner}},
  \bibinfo {author} {\bibfnamefont {G.}~\bibnamefont {Ehlers}}, \bibinfo
  {author} {\bibfnamefont {A.}~\bibnamefont {Fukaya}}, \bibinfo {author}
  {\bibfnamefont {E.}~\bibnamefont {Segal}}, \ and\ \bibinfo {author}
  {\bibfnamefont {Y.J.}\ \bibnamefont {Uemura}},\ }\bibfield  {title} {\enquote
  {\bibinfo {title} {Dynamic properties of a diluted pyrochlore cooperative
  paramagnet {(Tb$_p$Y$_{1-p}$)$_2$Ti$_2$O$_7$}},}\ }\href@noop {} {\bibfield
  {journal} {\bibinfo  {journal} {Phys. Rev. Lett.}\ }\textbf {\bibinfo
  {volume} {92}},\ \bibinfo {pages} {107204} (\bibinfo {year}
  {2004})}\BibitemShut {NoStop}%
\bibitem [{\citenamefont {Uemura}\ \emph {et~al.}(1985)\citenamefont {Uemura},
  \citenamefont {Yamazaki}, \citenamefont {Harshman}, \citenamefont {Senba},\
  and\ \citenamefont {Ansaldo}}]{uemura1985muon}%
  \BibitemOpen
  \bibfield  {author} {\bibinfo {author} {\bibfnamefont {Y.J.}\ \bibnamefont
  {Uemura}}, \bibinfo {author} {\bibfnamefont {T.}~\bibnamefont {Yamazaki}},
  \bibinfo {author} {\bibfnamefont {D.R.}\ \bibnamefont {Harshman}}, \bibinfo
  {author} {\bibfnamefont {M.}~\bibnamefont {Senba}}, \ and\ \bibinfo {author}
  {\bibfnamefont {E.J.}\ \bibnamefont {Ansaldo}},\ }\bibfield  {title}
  {\enquote {\bibinfo {title} {Muon-spin relaxation in {AuFe} and {CuMn} spin
  glasses},}\ }\href@noop {} {\bibfield  {journal} {\bibinfo  {journal} {Phys.
  Rev. B}\ }\textbf {\bibinfo {volume} {31}},\ \bibinfo {pages} {546} (\bibinfo
  {year} {1985})}\BibitemShut {NoStop}%
\bibitem [{\citenamefont {Uemura}\ \emph {et~al.}(1994)\citenamefont {Uemura},
  \citenamefont {Keren}, \citenamefont {Kojima}, \citenamefont {Le},
  \citenamefont {Luke}, \citenamefont {Wu}, \citenamefont {Ajiro},
  \citenamefont {Asano}, \citenamefont {Kuriyama}, \citenamefont {Mekata} \emph
  {et~al.}}]{uemura1994spin}%
  \BibitemOpen
  \bibfield  {author} {\bibinfo {author} {\bibfnamefont {Y.J.}\ \bibnamefont
  {Uemura}}, \bibinfo {author} {\bibfnamefont {A.}~\bibnamefont {Keren}},
  \bibinfo {author} {\bibfnamefont {K.}~\bibnamefont {Kojima}}, \bibinfo
  {author} {\bibfnamefont {L.P.}\ \bibnamefont {Le}}, \bibinfo {author}
  {\bibfnamefont {G.M.}\ \bibnamefont {Luke}}, \bibinfo {author} {\bibfnamefont
  {W.D.}\ \bibnamefont {Wu}}, \bibinfo {author} {\bibfnamefont
  {Y.}~\bibnamefont {Ajiro}}, \bibinfo {author} {\bibfnamefont
  {T.}~\bibnamefont {Asano}}, \bibinfo {author} {\bibfnamefont
  {Y.}~\bibnamefont {Kuriyama}}, \bibinfo {author} {\bibfnamefont
  {M.}~\bibnamefont {Mekata}},  \emph {et~al.},\ }\bibfield  {title} {\enquote
  {\bibinfo {title} {Spin fluctuations in frustrated kagom{\'e} lattice system
  {SrCr$_8$Ga$_4$O$_{19}$} studied by muon spin relaxation},}\ }\href@noop {}
  {\bibfield  {journal} {\bibinfo  {journal} {Phys. Rev. Lett.}\ }\textbf
  {\bibinfo {volume} {73}},\ \bibinfo {pages} {3306} (\bibinfo {year}
  {1994})}\BibitemShut {NoStop}%
\bibitem [{\citenamefont {Ogielski}(1985)}]{ogielski1985dynamics}%
  \BibitemOpen
  \bibfield  {author} {\bibinfo {author} {\bibfnamefont {A.T.}\ \bibnamefont
  {Ogielski}},\ }\bibfield  {title} {\enquote {\bibinfo {title} {Dynamics of
  three-dimensional ising spin glasses in thermal equilibrium},}\ }\href@noop
  {} {\bibfield  {journal} {\bibinfo  {journal} {Phys. Rev. B}\ }\textbf
  {\bibinfo {volume} {32}},\ \bibinfo {pages} {7384} (\bibinfo {year}
  {1985})}\BibitemShut {NoStop}%
\bibitem [{\citenamefont {Campbell}\ \emph {et~al.}(1994)\citenamefont
  {Campbell}, \citenamefont {Amato}, \citenamefont {Gygax}, \citenamefont
  {Herlach}, \citenamefont {Schenck}, \citenamefont {Cywinski},\ and\
  \citenamefont {Kilcoyne}}]{campbell1994dynamics}%
  \BibitemOpen
  \bibfield  {author} {\bibinfo {author} {\bibfnamefont {I.A.}\ \bibnamefont
  {Campbell}}, \bibinfo {author} {\bibfnamefont {A.}~\bibnamefont {Amato}},
  \bibinfo {author} {\bibfnamefont {F.N.}\ \bibnamefont {Gygax}}, \bibinfo
  {author} {\bibfnamefont {D.}~\bibnamefont {Herlach}}, \bibinfo {author}
  {\bibfnamefont {A.}~\bibnamefont {Schenck}}, \bibinfo {author} {\bibfnamefont
  {R.}~\bibnamefont {Cywinski}}, \ and\ \bibinfo {author} {\bibfnamefont
  {S.H.}\ \bibnamefont {Kilcoyne}},\ }\bibfield  {title} {\enquote {\bibinfo
  {title} {Dynamics in canonical spin glasses observed by muon spin
  depolarization},}\ }\href@noop {} {\bibfield  {journal} {\bibinfo  {journal}
  {Phys. Rev. Lett.}\ }\textbf {\bibinfo {volume} {72}},\ \bibinfo {pages}
  {1291} (\bibinfo {year} {1994})}\BibitemShut {NoStop}%
\bibitem [{\citenamefont {Keren}\ \emph {et~al.}(2001)\citenamefont {Keren},
  \citenamefont {Bazalitsky}, \citenamefont {Campbell},\ and\ \citenamefont
  {Lord}}]{keren2001probing}%
  \BibitemOpen
  \bibfield  {author} {\bibinfo {author} {\bibfnamefont {A.}~\bibnamefont
  {Keren}}, \bibinfo {author} {\bibfnamefont {G.}~\bibnamefont {Bazalitsky}},
  \bibinfo {author} {\bibfnamefont {I.}~\bibnamefont {Campbell}}, \ and\
  \bibinfo {author} {\bibfnamefont {J.S.}\ \bibnamefont {Lord}},\ }\bibfield
  {title} {\enquote {\bibinfo {title} {Probing exotic spin correlations by muon
  spin depolarization measurements with applications to spin glass dynamics},}\
  }\href@noop {} {\bibfield  {journal} {\bibinfo  {journal} {Phys. Rev. B}\
  }\textbf {\bibinfo {volume} {64}},\ \bibinfo {pages} {054403} (\bibinfo
  {year} {2001})}\BibitemShut {NoStop}%
\bibitem [{\citenamefont {Bono}\ \emph {et~al.}(2004)\citenamefont {Bono},
  \citenamefont {Mendels}, \citenamefont {Collin}, \citenamefont {Blanchard},
  \citenamefont {Bert}, \citenamefont {Amato}, \citenamefont {Baines},\ and\
  \citenamefont {Hillier}}]{bono2004mu}%
  \BibitemOpen
  \bibfield  {author} {\bibinfo {author} {\bibfnamefont {D.}~\bibnamefont
  {Bono}}, \bibinfo {author} {\bibfnamefont {P.}~\bibnamefont {Mendels}},
  \bibinfo {author} {\bibfnamefont {G.}~\bibnamefont {Collin}}, \bibinfo
  {author} {\bibfnamefont {N.}~\bibnamefont {Blanchard}}, \bibinfo {author}
  {\bibfnamefont {F.}~\bibnamefont {Bert}}, \bibinfo {author} {\bibfnamefont
  {A.}~\bibnamefont {Amato}}, \bibinfo {author} {\bibfnamefont
  {C.}~\bibnamefont {Baines}}, \ and\ \bibinfo {author} {\bibfnamefont {A.D.}\
  \bibnamefont {Hillier}},\ }\bibfield  {title} {\enquote {\bibinfo {title}
  {{$\mu$SR} study of the quantum dynamics in the frustrated $s=\frac32$
  kagom{\'e} bilayers},}\ }\href@noop {} {\bibfield  {journal} {\bibinfo
  {journal} {Phys. Rev. Lett.}\ }\textbf {\bibinfo {volume} {93}},\ \bibinfo
  {pages} {187201} (\bibinfo {year} {2004})}\BibitemShut {NoStop}%
\bibitem [{\citenamefont {Kermarrec}\ \emph {et~al.}(2011)\citenamefont
  {Kermarrec}, \citenamefont {Mendels}, \citenamefont {Bert}, \citenamefont
  {Colman}, \citenamefont {Wills}, \citenamefont {Strobel}, \citenamefont
  {Bonville}, \citenamefont {Hillier},\ and\ \citenamefont
  {Amato}}]{kermarrec2011spin}%
  \BibitemOpen
  \bibfield  {author} {\bibinfo {author} {\bibfnamefont {E.}~\bibnamefont
  {Kermarrec}}, \bibinfo {author} {\bibfnamefont {P.}~\bibnamefont {Mendels}},
  \bibinfo {author} {\bibfnamefont {F.}~\bibnamefont {Bert}}, \bibinfo {author}
  {\bibfnamefont {R.H.}\ \bibnamefont {Colman}}, \bibinfo {author}
  {\bibfnamefont {A.S.}\ \bibnamefont {Wills}}, \bibinfo {author}
  {\bibfnamefont {P.}~\bibnamefont {Strobel}}, \bibinfo {author} {\bibfnamefont
  {P.}~\bibnamefont {Bonville}}, \bibinfo {author} {\bibfnamefont
  {A.}~\bibnamefont {Hillier}}, \ and\ \bibinfo {author} {\bibfnamefont
  {A.}~\bibnamefont {Amato}},\ }\bibfield  {title} {\enquote {\bibinfo {title}
  {Spin-liquid ground state in the frustrated kagome antiferromagnet
  {MgCu$_3$(OH)$_6$Cl$_2$}},}\ }\href@noop {} {\bibfield  {journal} {\bibinfo
  {journal} {Phys. Rev. B}\ }\textbf {\bibinfo {volume} {84}},\ \bibinfo
  {pages} {100401} (\bibinfo {year} {2011})}\BibitemShut {NoStop}%
\bibitem [{\citenamefont {F{\aa}k}\ \emph {et~al.}(2012)\citenamefont
  {F{\aa}k}, \citenamefont {Kermarrec}, \citenamefont {Messio}, \citenamefont
  {Bernu}, \citenamefont {Lhuillier}, \citenamefont {Bert}, \citenamefont
  {Mendels}, \citenamefont {Koteswararao}, \citenamefont {Bouquet},
  \citenamefont {Ollivier} \emph {et~al.}}]{faak2012kapellasite}%
  \BibitemOpen
  \bibfield  {author} {\bibinfo {author} {\bibfnamefont {B.}~\bibnamefont
  {F{\aa}k}}, \bibinfo {author} {\bibfnamefont {E.}~\bibnamefont {Kermarrec}},
  \bibinfo {author} {\bibfnamefont {L.}~\bibnamefont {Messio}}, \bibinfo
  {author} {\bibfnamefont {B.}~\bibnamefont {Bernu}}, \bibinfo {author}
  {\bibfnamefont {C.}~\bibnamefont {Lhuillier}}, \bibinfo {author}
  {\bibfnamefont {F.}~\bibnamefont {Bert}}, \bibinfo {author} {\bibfnamefont
  {P.}~\bibnamefont {Mendels}}, \bibinfo {author} {\bibfnamefont
  {B.}~\bibnamefont {Koteswararao}}, \bibinfo {author} {\bibfnamefont
  {F.}~\bibnamefont {Bouquet}}, \bibinfo {author} {\bibfnamefont
  {J.}~\bibnamefont {Ollivier}},  \emph {et~al.},\ }\bibfield  {title}
  {\enquote {\bibinfo {title} {Kapellasite: A kagome quantum spin liquid with
  competing interactions},}\ }\href@noop {} {\bibfield  {journal} {\bibinfo
  {journal} {Phys. Rev. Lett.}\ }\textbf {\bibinfo {volume} {109}},\ \bibinfo
  {pages} {037208} (\bibinfo {year} {2012})}\BibitemShut {NoStop}%
\bibitem [{\citenamefont {Clark}\ \emph {et~al.}(2013)\citenamefont {Clark},
  \citenamefont {Orain}, \citenamefont {Bert}, \citenamefont {{De Vries}},
  \citenamefont {Aidoudi}, \citenamefont {Morris}, \citenamefont {Lightfoot},
  \citenamefont {Lord}, \citenamefont {Telling}, \citenamefont {Bonville} \emph
  {et~al.}}]{clark2013gapless}%
  \BibitemOpen
  \bibfield  {author} {\bibinfo {author} {\bibfnamefont {L.}~\bibnamefont
  {Clark}}, \bibinfo {author} {\bibfnamefont {J.C.}\ \bibnamefont {Orain}},
  \bibinfo {author} {\bibfnamefont {F.}~\bibnamefont {Bert}}, \bibinfo {author}
  {\bibfnamefont {M.A.}\ \bibnamefont {{De Vries}}}, \bibinfo {author}
  {\bibfnamefont {F.H.}\ \bibnamefont {Aidoudi}}, \bibinfo {author}
  {\bibfnamefont {R.~E.}\ \bibnamefont {Morris}}, \bibinfo {author}
  {\bibfnamefont {P.}~\bibnamefont {Lightfoot}}, \bibinfo {author}
  {\bibfnamefont {J.S.}\ \bibnamefont {Lord}}, \bibinfo {author} {\bibfnamefont
  {M.T.F.}\ \bibnamefont {Telling}}, \bibinfo {author} {\bibfnamefont
  {P.}~\bibnamefont {Bonville}},  \emph {et~al.},\ }\bibfield  {title}
  {\enquote {\bibinfo {title} {Gapless spin liquid ground state in the
  $s=\frac12$ vanadium oxyfluoride kagome antiferromagnet {[NH$_4]_2$[C$_7$
  H$_{14}$N][V$_7$O$_6$F$_{18}$]}},}\ }\href@noop {} {\bibfield  {journal}
  {\bibinfo  {journal} {Phys. Rev. Lett.}\ }\textbf {\bibinfo {volume} {110}},\
  \bibinfo {pages} {207208} (\bibinfo {year} {2013})}\BibitemShut {NoStop}%
\bibitem [{\citenamefont {Gomilsek}\ \emph {et~al.}()\citenamefont {Gomilsek},
  \citenamefont {Klanjsek}, \citenamefont {Pregelj}, \citenamefont {Luetkens},
  \citenamefont {Li}, \citenamefont {Zhang},\ and\ \citenamefont
  {Zorko}}]{gomilsek2016mu}%
  \BibitemOpen
  \bibfield  {author} {\bibinfo {author} {\bibfnamefont {M.}~\bibnamefont
  {Gomilsek}}, \bibinfo {author} {\bibfnamefont {M.}~\bibnamefont {Klanjsek}},
  \bibinfo {author} {\bibfnamefont {M.}~\bibnamefont {Pregelj}}, \bibinfo
  {author} {\bibfnamefont {H.}~\bibnamefont {Luetkens}}, \bibinfo {author}
  {\bibfnamefont {Y.}~\bibnamefont {Li}}, \bibinfo {author} {\bibfnamefont
  {Q.M.}\ \bibnamefont {Zhang}}, \ and\ \bibinfo {author} {\bibfnamefont
  {A.}~\bibnamefont {Zorko}},\ }\href@noop {} {\enquote {\bibinfo {title}
  {{$\mu$SR} insight into the impurity problem in quantum kagome
  antiferromagnets},}\ }\bibinfo {note} {{a}rXiv:1604.02916}\BibitemShut
  {NoStop}%
\bibitem [{\citenamefont {Gomil{\v{s}}ek}\ \emph {et~al.}(2016)\citenamefont
  {Gomil{\v{s}}ek}, \citenamefont {Klanj{\v{s}}ek}, \citenamefont {Pregelj},
  \citenamefont {Coomer}, \citenamefont {Luetkens}, \citenamefont {Zaharko},
  \citenamefont {Fennell}, \citenamefont {Li}, \citenamefont {Zhang},\ and\
  \citenamefont {Zorko}}]{gomilvsek2016instabilities}%
  \BibitemOpen
  \bibfield  {author} {\bibinfo {author} {\bibfnamefont {M.}~\bibnamefont
  {Gomil{\v{s}}ek}}, \bibinfo {author} {\bibfnamefont {M.}~\bibnamefont
  {Klanj{\v{s}}ek}}, \bibinfo {author} {\bibfnamefont {M.}~\bibnamefont
  {Pregelj}}, \bibinfo {author} {\bibfnamefont {F.C.}\ \bibnamefont {Coomer}},
  \bibinfo {author} {\bibfnamefont {H.}~\bibnamefont {Luetkens}}, \bibinfo
  {author} {\bibfnamefont {O.}~\bibnamefont {Zaharko}}, \bibinfo {author}
  {\bibfnamefont {T.}~\bibnamefont {Fennell}}, \bibinfo {author} {\bibfnamefont
  {Y.}~\bibnamefont {Li}}, \bibinfo {author} {\bibfnamefont {Q.M.}\
  \bibnamefont {Zhang}}, \ and\ \bibinfo {author} {\bibfnamefont
  {A.}~\bibnamefont {Zorko}},\ }\bibfield  {title} {\enquote {\bibinfo {title}
  {Instabilities of spin-liquid states in a quantum kagome antiferromagnet},}\
  }\href@noop {} {\bibfield  {journal} {\bibinfo  {journal} {Phys. Rev. B}\
  }\textbf {\bibinfo {volume} {93}},\ \bibinfo {pages} {060405} (\bibinfo
  {year} {2016})}\BibitemShut {NoStop}%
\bibitem [{\citenamefont {Avella}\ and\ \citenamefont
  {Mancini}(2013)}]{avella2013strongly}%
  \BibitemOpen
  \bibfield  {author} {\bibinfo {author} {\bibfnamefont {A.}~\bibnamefont
  {Avella}}\ and\ \bibinfo {author} {\bibfnamefont {F.}~\bibnamefont
  {Mancini}},\ }\href@noop {} {\emph {\bibinfo {title} {Strongly correlated
  systems: numerical methods}}},\ Vol.\ \bibinfo {volume} {176}\ (\bibinfo
  {publisher} {Springer Science \& Business Media},\ \bibinfo {year}
  {2013})\BibitemShut {NoStop}%
\bibitem [{\citenamefont {Hohenberg}\ and\ \citenamefont
  {Halperin}(1977)}]{hohenberg1977theory}%
  \BibitemOpen
  \bibfield  {author} {\bibinfo {author} {\bibfnamefont {P.C.}\ \bibnamefont
  {Hohenberg}}\ and\ \bibinfo {author} {\bibfnamefont {B.I.}\ \bibnamefont
  {Halperin}},\ }\bibfield  {title} {\enquote {\bibinfo {title} {Theory of
  dynamic critical phenomena},}\ }\href@noop {} {\bibfield  {journal} {\bibinfo
   {journal} {Rev. Mod. Phys.}\ }\textbf {\bibinfo {volume} {49}},\ \bibinfo
  {pages} {435} (\bibinfo {year} {1977})}\BibitemShut {NoStop}%
\bibitem [{\citenamefont {McMullen}\ and\ \citenamefont
  {Zaremba}(1978)}]{mcmullen1978positive}%
  \BibitemOpen
  \bibfield  {author} {\bibinfo {author} {\bibfnamefont {T.}~\bibnamefont
  {McMullen}}\ and\ \bibinfo {author} {\bibfnamefont {E.}~\bibnamefont
  {Zaremba}},\ }\bibfield  {title} {\enquote {\bibinfo {title} {Positive-muon
  spin depolarization in solids},}\ }\href@noop {} {\bibfield  {journal}
  {\bibinfo  {journal} {Phys. Rev. B}\ }\textbf {\bibinfo {volume} {18}},\
  \bibinfo {pages} {3026} (\bibinfo {year} {1978})}\BibitemShut {NoStop}%
\bibitem [{\citenamefont {Hermele}\ \emph {et~al.}(2004)\citenamefont
  {Hermele}, \citenamefont {Fisher},\ and\ \citenamefont
  {Balents}}]{hermele2004}%
  \BibitemOpen
  \bibfield  {author} {\bibinfo {author} {\bibfnamefont {M.}~\bibnamefont
  {Hermele}}, \bibinfo {author} {\bibfnamefont {M.~P.~A.}\ \bibnamefont
  {Fisher}}, \ and\ \bibinfo {author} {\bibfnamefont {L.}~\bibnamefont
  {Balents}},\ }\bibfield  {title} {\enquote {\bibinfo {title} {Pyrochlore
  photons: The {$U(1)$} spin liquid in a $s=\frac12$ three-dimensional
  frustrated magnet},}\ }\href@noop {} {\bibfield  {journal} {\bibinfo
  {journal} {Phys. Rev. B}\ }\textbf {\bibinfo {volume} {69}},\ \bibinfo
  {pages} {064404} (\bibinfo {year} {2004})}\BibitemShut {NoStop}%
\bibitem [{\citenamefont {Savary}\ and\ \citenamefont
  {Balents}(2012)}]{savary2012}%
  \BibitemOpen
  \bibfield  {author} {\bibinfo {author} {\bibfnamefont {L.}~\bibnamefont
  {Savary}}\ and\ \bibinfo {author} {\bibfnamefont {L.}~\bibnamefont
  {Balents}},\ }\bibfield  {title} {\enquote {\bibinfo {title} {Coulombic
  quantum liquids in spin-1/2 pyrochlores},}\ }\href@noop {} {\bibfield
  {journal} {\bibinfo  {journal} {Phys. Rev. Lett.}\ }\textbf {\bibinfo
  {volume} {108}},\ \bibinfo {pages} {037202} (\bibinfo {year}
  {2012})}\BibitemShut {NoStop}%
\bibitem [{\citenamefont {Lee}(2008{\natexlab{b}})}]{lee2008}%
  \BibitemOpen
  \bibfield  {author} {\bibinfo {author} {\bibfnamefont {P.~A.}\ \bibnamefont
  {Lee}},\ }\bibfield  {title} {\enquote {\bibinfo {title} {From high
  temperature superconductivity to quantum spin liquid: progress in strong
  correlation physics},}\ }\href@noop {} {\bibfield  {journal} {\bibinfo
  {journal} {Rep. Prog. Phys.}\ }\textbf {\bibinfo {volume} {71}},\ \bibinfo
  {pages} {012501} (\bibinfo {year} {2008}{\natexlab{b}})},\ \bibinfo {note}
  {arXiv:0708.2115}\BibitemShut {NoStop}%
\bibitem [{\citenamefont {Nave}\ and\ \citenamefont {Lee}(2007)}]{nave2007}%
  \BibitemOpen
  \bibfield  {author} {\bibinfo {author} {\bibfnamefont {C.~P.}\ \bibnamefont
  {Nave}}\ and\ \bibinfo {author} {\bibfnamefont {P.~A.}\ \bibnamefont {Lee}},\
  }\bibfield  {title} {\enquote {\bibinfo {title} {Transport properties of a
  spinon {Fermi} surface coupled to a {$U(1)$} gauge field},}\ }\href@noop {}
  {\bibfield  {journal} {\bibinfo  {journal} {Phys. Rev. B}\ }\textbf {\bibinfo
  {volume} {76}},\ \bibinfo {pages} {235124} (\bibinfo {year}
  {2007})}\BibitemShut {NoStop}%
\bibitem [{\citenamefont {Ng}\ and\ \citenamefont {Lee}(2007)}]{ng2007}%
  \BibitemOpen
  \bibfield  {author} {\bibinfo {author} {\bibfnamefont {T.-K.}\ \bibnamefont
  {Ng}}\ and\ \bibinfo {author} {\bibfnamefont {P.~A.}\ \bibnamefont {Lee}},\
  }\bibfield  {title} {\enquote {\bibinfo {title} {Power-law conductivity
  inside the {Mott} gap: Application to
  {$\kappa$-(BEDT-TTF)$_2$Cu$_2$(CN)$_3$}},}\ }\href@noop {} {\bibfield
  {journal} {\bibinfo  {journal} {Phys. Rev. Lett.}\ }\textbf {\bibinfo
  {volume} {99}},\ \bibinfo {pages} {156402} (\bibinfo {year}
  {2007})}\BibitemShut {NoStop}%
\bibitem [{\citenamefont {Ma}\ and\ \citenamefont {Ng}(2015)}]{ma2015}%
  \BibitemOpen
  \bibfield  {author} {\bibinfo {author} {\bibfnamefont {Y.-F.}\ \bibnamefont
  {Ma}}\ and\ \bibinfo {author} {\bibfnamefont {T.-K.}\ \bibnamefont {Ng}},\
  }\bibfield  {title} {\enquote {\bibinfo {title} {Surface plasmons and
  reflectance of {$U(1)$} spin liquid states with large spinon {Fermi}
  surfaces},}\ }\href@noop {} {\bibfield  {journal} {\bibinfo  {journal} {Phys.
  Rev. B}\ }\textbf {\bibinfo {volume} {91}},\ \bibinfo {pages} {075106}
  (\bibinfo {year} {2015})}\BibitemShut {NoStop}%
\bibitem [{\citenamefont {Li}\ \emph {et~al.}()\citenamefont {Li},
  \citenamefont {Wang},\ and\ \citenamefont {Chen}}]{li2015}%
  \BibitemOpen
  \bibfield  {author} {\bibinfo {author} {\bibfnamefont {Y.-D.}\ \bibnamefont
  {Li}}, \bibinfo {author} {\bibfnamefont {X.}~\bibnamefont {Wang}}, \ and\
  \bibinfo {author} {\bibfnamefont {G.}~\bibnamefont {Chen}},\ }\href@noop {}
  {\enquote {\bibinfo {title} {An anisotropic spin model of strong
  spin-orbit-coupled triangular antiferromagnets},}\ }\bibinfo {note}
  {{arXiv:1512.02151}}\BibitemShut {NoStop}%
\end{thebibliography}%

\end{document}